\documentclass[aps,prb,twocolumn,showpacs,superscriptaddress,amsmath,amssymb]{revtex4-1}

\usepackage{siunitx}
\usepackage{setspace}
\usepackage{graphicx}
\usepackage{amssymb}
\usepackage{xcolor}
\usepackage{subfig}
\usepackage[section]{placeins}
\usepackage{hyperref}
\usepackage[capitalise]{cleveref}
\usepackage[version=3]{mhchem}

\begin{document}

\title{Excitonic recombinations in hBN: from bulk to exfoliated layers}

\author{A. Pierret}
\affiliation{Laboratoire d'Etude des Microstructures, ONERA-CNRS, BP 72, 92322 Ch\^atillon Cedex, France}
\affiliation{CEA-CNRS Group ``Nanophysique et Semiconducteurs'', Institut N\'eel/CNRS-Universit\'e J. Fourier and CEA Grenoble, INAC, SP2M, 17 rue des Martyrs, 38054 Grenoble Cedex 9, France}

\author{J. Loayza}

\affiliation{Laboratoire d'Etude des Microstructures, ONERA-CNRS, BP 72, 92322 Ch\^atillon Cedex, France}
\affiliation{Groupe d'Etude de la Mati\`ere Condens\'ee, University of Versailles St-Quentin and CNRS, 45 avenue des Etats-Unis, 78000 Versailles, France}

\author{B. Berini}
\affiliation{Groupe d'Etude de la Mati\`ere Condens\'ee, University of Versailles St-Quentin and CNRS, 45 avenue des Etats-Unis, 78000 Versailles, France}

\author{A. Betz}
\altaffiliation{Current address: Hitachi Cambridge Laboratory, J. J. Thomson Avenue, Cambridge CB23 7GP, United Kingdom}
\author{B. Pla\c{c}ais}
\affiliation{Laboratoire Pierre Aigrain, ENS-CNRS UMR 8551, Universit\'es P. et M. Curie and Paris-Diderot, 24, rue Lhomond, 75231 Paris Cedex 05, France}

\author{F. Ducastelle}
\affiliation{Laboratoire d'Etude des Microstructures, ONERA-CNRS, BP 72, 92322 Ch\^atillon Cedex, France}

\author{J. Barjon}
\email{julien.barjon@uvsq.fr}
\affiliation{Groupe d'Etude de la Mati\`ere Condens\'ee, University of Versailles St-Quentin and CNRS, 45 avenue des Etats-Unis, 78000 Versailles, France}

\author{A. Loiseau}
\email{annick.loiseau@onera.fr}
\affiliation{Laboratoire d'Etude des Microstructures, ONERA-CNRS, BP 72, 92322 Ch\^atillon Cedex, France}

\begin{abstract}
Hexagonal boron nitride ($h$-BN) and graphite are structurally similar but with very different properties. Their combination in graphene-based devices is now of intense research focus, and it becomes particularly important to evaluate the role played by crystalline defects on their properties. In this paper, the cathodoluminescence (CL) properties of hexagonal boron nitride crystallites are reported and compared to those of nanosheets mechanically exfoliated from them. First the link between the presence of structural defects and the recombination intensity of trapped excitons, the so-called $D$ series, is confirmed. Low defective $h$-BN regions are further evidenced by CL spectral mapping (hyperspectral imaging), allowing us to observe new features in the near-band-edge region, tentatively attributed to phonon replica of exciton recombinations. Second the $h$-BN thickness was reduced down to six atomic layers, using mechanical exfoliation, as evidenced by atomic force microscopy. Even at these low thicknesses, the luminescence remains intense and exciton recombination energies are not strongly modified with respect to the bulk, as expected from theoretical calculations indicating extremely compact excitons in $h$-BN.
\end{abstract}


\maketitle

\section{Introduction}

Hexagonal boron nitride ($h$-BN) has the same honeycomb lattice as graphite with two atoms per unit cell and similar lattice parameters. Due to this similarity, boron nitride materials have attracted  a growing interest in line with the development of low-dimensional carbon-related materials. Similarly to  carbon, BN materials can be synthesized either as nanotubes (one-dimensional (1D) form) \cite{Chopra1995,Loiseau1996} or as monolayers and/or multilayers (two-dimensional (2D) form).\cite{Shi10,Song10} In the following we focus on this latter form. 2D layers of carbon, namely graphene sheets, display extraordinary electronic properties which open unanticipated routes for a new generation of electronic devices. However, the electron mobility of supported graphene typically falls short of that of suspended graphene, due to detrimental effects of substrate disorder and adsorbents. \cite{Burson13,Xue11,Decker11} Facing this problem, $h$-BN layers are of particular interest as support or capping layers of graphene. They  combine several properties: They are insulating ($h$-BN is a large gap semiconductor due to the polar BN bond), \cite{Zunger1976,Museur11,Blase1995} they display an especially compatible layered $sp^2$ structure with that of graphene, they have a low concentration of charges impurities and they can be very flat due to an easy cleavage. Owing to these properties, graphene transferred on BN layers displays an electron mobility at room temperature of \SI{120 000}{cm^2/V s}, which is the highest reported value for a supported graphene \cite{Dean10,Gannett11,Mayorov11,Zomer11} and very close to that of suspended graphene.  \cite{Bolotin2008,Du2008} Beyond the high mobility of graphene supported on BN,  their excellent lattice matching is promising for the realization of heterostructures of these materials for vertical transport stacking, in which graphene layers act as tunable metallic electrodes for the BN quasi-ideal tunnel barrier. \cite{Britnell12,Ramasubramaniam11}
These promising perspectives have been  demonstrated by pioneering experiments done using  sheets mechanically exfoliated from both graphite and $h$-BN single crystals. In the future, $h$-BN and graphene based devices and heterostructures would most probably use chemical vapor deposited (CVD)  polycrystalline films and sheets. Their performances would only be achieved via an accurate control of the defects in both graphene and BN layers and of the layers engineering. While the electronic properties of graphene have been well described theoretically and investigated experimentally, this is not the case of BN layers and even of $h$-BN. This is due to both the scarcity of high quality materials and to the nature of their electronic properties dictated by the large gap. It is thus a basic issue to understand the spectroscopic properties of atomically thin $h$-BN layers and their intrinsic defects, which is the focus of this paper.

In contrast to graphene, usual spectroscopic characterization techniques such as Raman are not easy to manipulate or they provide  poor information when used for $h$-BN. Absorption and luminescence spectroscopies have been shown to be the most direct approach to investigate the electronic properties of BN materials, due to their large gap. To this aim, dedicated cathodolumnescence and photoluminescence experiments have been recently developed and applied to BN powders and single crystals. \cite{JaffrennouJAP2007,WatanabePRB2009,MuseurJAP2008,Museur11} Both theoretical calculations \cite{WirtzArXiv2005,Arnaud2006,WirtzComment2008,ArnaudReply2008} and the most recent excitation photoluminescence experiments on single crystals \cite{Museur11} converge to establish the band gap of $h$-BN near \SI{6.4}{eV}. Furthermore, it is now commonly accepted that $h$-BN optical properties are dominated by huge excitonic effects. The near-band-edge luminescence spectrum is composed of two series of lines. Referring to measurements done on single crystals in Ref. [\onlinecite{WatanabePRB2009}], they are defined as the $S$ and $D$ series. The four higher energy lines, labeled $S1$ to $S4$, located between  5.7 and \SI{5.9}{eV}, are attributed to the excitons, whereas the lower energy ones, labeled $D1$ to $D4$, between 5.4 and  \SI{5.65}{eV}, are assigned to excitons trapped to structural defects. \cite{WatanabePRB2009,JaffrennouJAP2007}
The excitons in $h$-BN are more of Frenkel-type than of Wannier-type (as in others usual semiconductors, such as AlN with a \SI{6.1}{eV} gap). \textit{Ab initio} calculations indeed predict that the spatial extension of the exciton wavefunction is of the order of one $h$-BN atomic layer.\cite{WirtzArXiv2005,Arnaud2006,ArnaudReply2008,WirtzComment2008} Moreover the experimental Stokes shift of  \SI{250}{meV} observed for the $S4$-line suggests its self-trapping, \cite{Museur11,WatanabePRB2009} consistent with the very localized view of the Frenkel exciton.

To complete this view, the effect of a reduction in the $h$-BN thickness down to the atomic level has to be analyzed. Up to now, only scarce studies deal with the optical properties of nanometer-thick BN layers. An optical absorption edge between 5.6 and \SI{6.1}{eV} at room temperature is reported,  \cite{Kim12,WangBN12,Ismach12,PakdelJMC12,Ci10,Shi10,Song10,Lin2009,Gao12}  \textit{i.e.} in the same range than in bulk $h$-BN. Only two studies report near-band edge recombination luminescence, with no correlation to the BN layer thickness under investigation. \cite{Li12,WangBN12}

In this paper we present the first study of the luminescence properties of single BN nanosheets, with well-known thickness, by combining atomic force microscopy (AFM) and cathodoluminescence (CL) measurements. BN nanosheets were prepared by mechanical exfoliation of small $h$-BN crystallites of a polycrystalline powder. This material offers the advantage to give access at the same time to the intrinsic optical response of the crystallite as well as to the effect of grain boundaries and the crystallite thickness on this response. An advanced characterization of the starting bulk material is first presented and its near-band-edge recombinations observed by CL are discussed with respect to those of the single crystal. Then the luminescence of the exfoliated BN sheets is presented and discussed as a function of their thickness.

\section{Experimental}

\subsection{Samples and exfoliation process}

The bulk material exfoliated in this paper is the high purity TR\`ES BN$^{\copyright}$ St-Gobain commercial power PUHP1108, used in cosmetic applications. The $h$-BN crystallites are already shaped as flakes with large (00.1) surfaces of typically  \SI{10}{\micro\meter} diameter, which is particularly convenient for the exfoliation process. Their thickness is about \SI{100}{nm}. One of these crystallites is shown in the scanning electron microscope (SEM) image of  \Cref{SEM}. This powder was synthesized at high temperature from boric acid and a nitrogen source. Our measurements on this powder are compared with the ones obtained from a single crystal of the best available quality, synthesized by a high-pressure high-temperature (HPHT) crystal process \cite{Taniguchi2007} and provided by Taniguchi \textit{et al.} This reference sample is  hereafter referred to as the HPHT sample.

Exfoliation of few-layer $h$-BN was carried out by mechanical peeling following the same method used for graphene. \cite{Novoselov2004} The powder is applied to an adhesive tape, whose repeated folding and peeling apart separates the layers. These thin layers are then transferred on a Si wafer covered with \SI{90}{nm} of \ce{SiO2}, which is the optimal thickness for imaging BN flakes with  maximum contrast.  \cite{Gorbachev11} Prior to the $h$-BN deposition, Cr/Au localization marks were formed on the wafer by means of UV-lithography using an AZ5214E photo-resist and Joule evaporation. \cite{BetzPhD} The marks facilitate the localization of the flakes for the different measurements to be achieved on a given flake. Still prior to transferring the layers, the wafer is chemically cleaned with acetone and isopropanol, followed by several minutes of exposure to a strong  \ce{O2} plasma (\SI{60}{\watt}, $P\leq \SI{12}{\nano\bar}$). This last step eliminates most contaminants but also renders the  \ce{SiO2} surface hydrophilic,\cite{Nagashio11}  which has the drawback of facilitating the inclusion of a thin water layer between the flakes and the substrate.\cite{Ishigami2007,Nagashio11}  As we shall see later, this complicates the layer thickness measurement done by atomic force microscopy (AFM). The surface morphology of the $h$-BN exfoliated layers was investigated by atomic force microscopy with a Dimension 3100 scanning probe microscope (Brukers) operating in tapping mode with commercial MPP-11100 probes.

\subsection{Cathodoluminescence procedure}

The cathodoluminescence of the $h$-BN samples was analyzed at low temperature in the (i) spectroscopic, (ii) imaging and (iii) spectral mapping modes, using an optical system (Horiba Jobin Yvon SA) installed on a high resolution JEOL7001F field-emission scanning electron microscope. The samples are mounted on a GATAN cryostat SEM stage and cooled down to $\approx \SI{20}{\kelvin}$ with a continuous flow of liquid helium. They are excited by electrons accelerated at  \SI{5}{kV} with a beam current as low as  \SI{0.17}{nA}. The CL emission is collected by a parabolic mirror and focused with mirror optics on the entrance slit of a \SI{55}{cm}-focal length monochromator. The all-mirror optics combined with a suitable choice of UV detectors and gratings ensures a high spectral sensitivity down to  \SI{190}{nm}.  A silicon charge-coupled-display (CCD) camera is used to record spectra in mode (i) and as well for the spectral mapping mode (iii). In mode (iii), also referred as the hyperspectral imaging acquisition mode, the focused electron beam is scanned step by step with the  \textit{HJY CL Link} drive unit (co-developed and tested at GEMaC) and synchronized with the CCD camera to record one spectrum for each point. The spectrometer is also equipped with a UV photomultiplier on the lateral side exit for fast monochromatic CL imaging (\textit{i.e.} image of the luminescence at a given wavelength) in mode (ii). For all spectra reported in this paper, it was checked that the linewidths are not limited by the spectral resolution of the apparatus (\SI{0.02}{nm} in the best conditions). 
In the case of $h$-BN exfoliated layers, CL was only performed in mode (i) and using a fast \textit{e}-beam scanning of a 1.5$\times$\SI{1.2}{\micro\meter\squared} area on the sample instead of using a fixed focused beam, in order to minimize the irradiation dose and the \textit{e}-beam induced modifications.

\section{Results and discussion}

\subsection{Luminescence of the bulk material}

\begin{figure}
	\centering
	\subfloat{
		\includegraphics[scale=0.55]{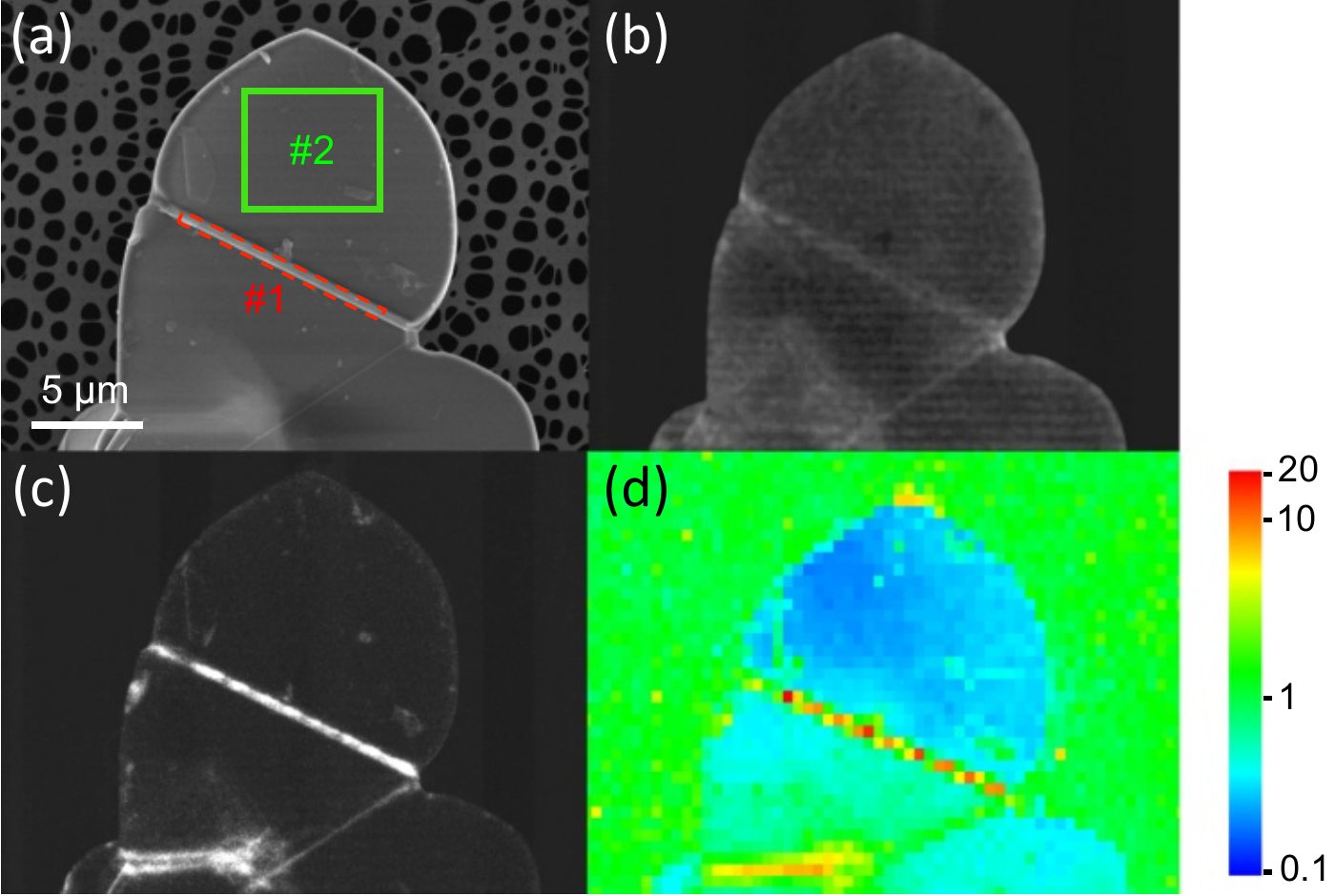}\label{SEM}}
	\subfloat{
		\label{CL_S}}
	\subfloat{
		\label{CL_D}}
	\subfloat{
		\label{CL_D/S}}
	\caption{Images of the bulk material: (a) SEM image of the analyzed crystallite; (b) (c) Corresponding monochromatic CL images recorded at (b) \SI{5.76}{eV} (\SI{215}{nm}), centered on the main $S$ line emission and at (c) \SI{5.46}{eV} (\SI{227}{nm}), centered on the main $D$ line emission; (d) Map of the $D/S$ ratio between the structural defect-related and the intrinsic excitonic recombinations, extracted from a 64x48 CL hyperspectral-mapping.}
	\label{cryst}
\end{figure}

The $h$-BN crystallites are often made of a few single crystals separated by grain boundaries, as for another powder studied in a previous work.\cite{JaffrennouJAP2007}  Owing to the electron diffraction patterns recorded by transmission electron microscopy (TEM) on the specimen presented in  \Cref{SEM}, each grain could be precisely orientated along the (00.1) zone axis (not shown here, for experimental details, see Ref. [\onlinecite{JaffrennouJAP2007}]). The TEM analysis shows that the two main grains of the crystallite in  \Cref{SEM} are slightly tilted one with respect to the other. They are separated by the grain boundary labeled  \#1.

\begin{figure}
	\centering
	\includegraphics{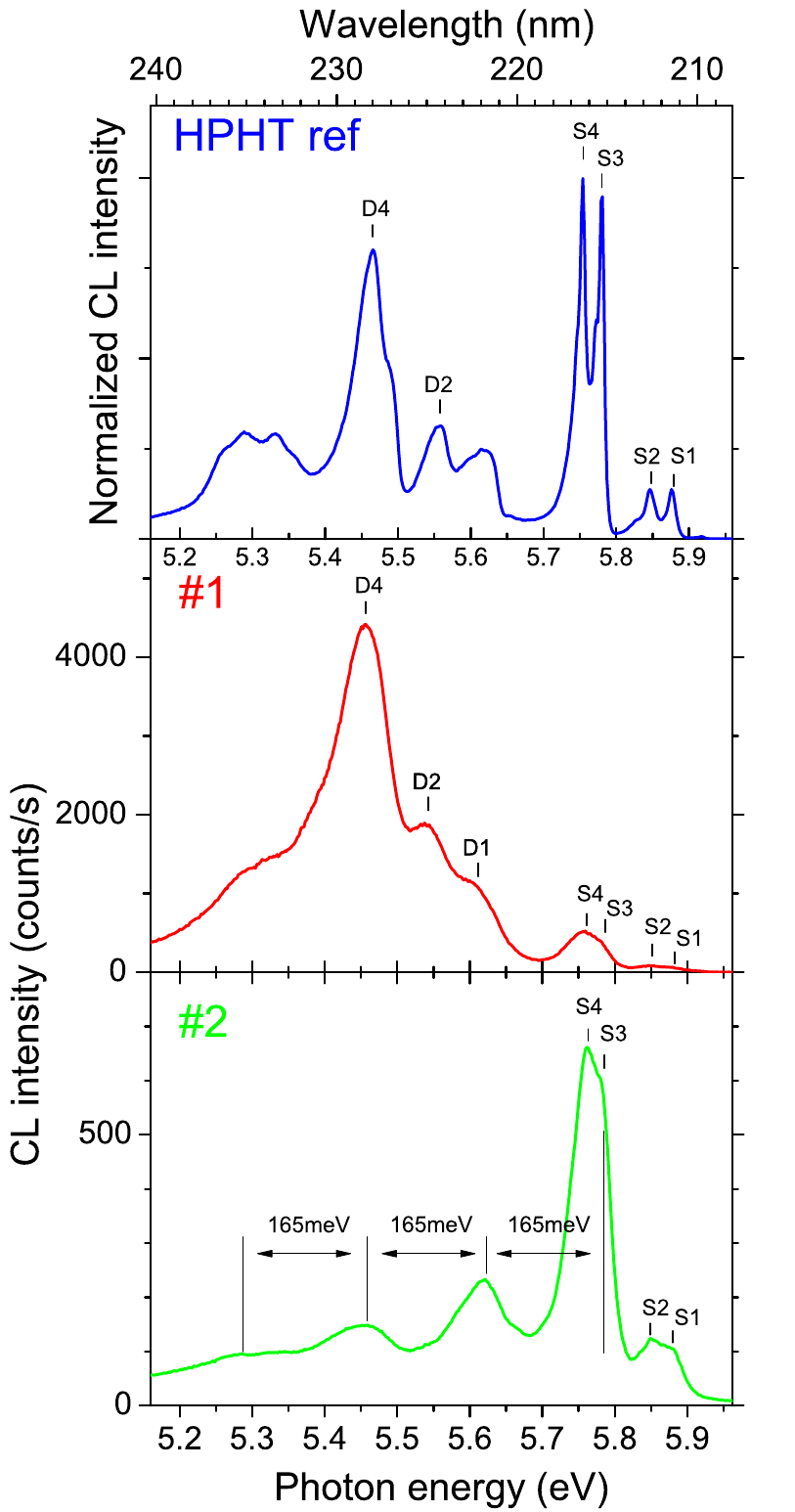}
	\caption{CL spectra of the bulk  $h$-BN materials in the near-band edge region: at the top is a reference spectrum of a HPHT high quality crystal,  \cite{Taniguchi2007}  compared to \#1, registered in the grain boundary area delimited by the red rectangle (averaged over 15 spectra) and to \#2 on the main grain of the crystallite in the area delimited by the green square (averaged over 240 spectra), as labeled in \Cref{SEM}. The specimen temperature is about \SI{20}{\kelvin}.}
	\label{cryst_spectra}
\end{figure}

CL spectra recorded on different areas (labelled \#1 and \#2) of the crystallite in \Cref{SEM} are shown in \Cref{cryst_spectra}. They are compared to the one recorded on the HPHT reference sample and  hereafter referred to as the reference spectrum. It is worth mentioning that this reference sample precisely displays the same exciton recombination energies than the ones observed in photoluminescence experiments done on a single crystal of the same origin as reported in Ref. [\onlinecite{WatanabePRB2009}].

Spectrum \#1 in \Cref{cryst_spectra} is recorded at the grain boundary of the $h$-BN crystallite. It presents the same features as the reference sample, with clear $S$and $D$ series. The $D$ lines are slightly shifted toward lower energies by about \SI{15}{meV} when compared to the reference spectrum. Since the S lines are not shifted, such a decrease of the emission energy is attributed to a different binding energy of the exciton to the grain boundary, rather than to a change of bandgap energy. Moreover the $S$ and $D$ series appear broader than in the reference sample, possibly due to an inhomogeneous distribution of residual strain in the sample. This interpretation is supported by the slight fluctuations of the recombination energies of a few meV, observed from point to point in a CL mapping (not shown here).

Moreover we remark that the relative intensities of $S$ and $D$ series are very different when compared to the reference spectrum. Spectrum \#1 is dominated by the $D$ series. It is known that this series is linked to the presence of structural defects such as grain boundaries, dislocations, or stacking faults. \cite{JaffrennouJAP2007,WatanabePRB2009,WatanabeAPL2006} To confirm this interpretation, monochromatic images have been registered, at energy centered on the $S4$ line (at \SI{5.76}{eV}, \Cref{CL_S}) and on the D4 series (at \SI{5.46}{eV}, \Cref{CL_D}). Both $D$ and $S$ emission series are slightly enhanced (by a factor of two in average) at the crystallite edges. This enhancement confirms previous observations on single $h$-BN crystals by Kubota \textit{et al.}, \cite{Kubota2007} and is consistent with their interpretation based on a more efficient light scattering at the grain edges. More interestingly, a clear one-to-one correlation is observed between the $D$ series emission and the grain boundary locations (\Cref{SEM,CL_D}).

To better quantify this view and to dispose of a comparison criterion, we define the $D/S$ ratio as the ratio of amplitudes between the $D4$ and the $S4$ peaks (at  \SI{5.46}{eV} and \SI{5.77}{eV}, resp. 227 and \SI{215}{nm}). It can be viewed as an indicator of the defect density, as introduced by Watanabe \textit{et al.} \cite{WatanabeAPL2006} and in the same way than in conventional semiconductors for the impurities content. \cite{Barjon11}

Owing to CL spectral mapping facilities, the $D/S$ ratio is evaluated in each point of the studied $h$-BN crystallite of \Cref{SEM}, and shown in \Cref{CL_D/S}. It exhibits a variation of two orders of magnitude (0.2 -- 20). This reveals the extremely high impact of structural defects in bulk $h$-BN.

It is also remarkable that in the central part of the upper grain, the $D/S$ ratio is four times lower than in our reference HPHT sample ($D/S>0.8$), suggesting a better structural quality. As the $D$ series almost vanishes, new intrinsic features are revealed, which were hindered by the $D$ series in previous studies.  Indeed in the CL spectrum \#2 of \Cref{cryst_spectra} recorded from the central part of the grain, three new lines are detected at 5.619, 5.457 and \SI{5.285}{eV}. They are shifted by $n$ (with $n = 1, 2, 3$) times \SI{165}{meV} (\SI{1330}{\per\cm}) from the $S3-4$ band, a period close to the  \SI{169}{meV} (\SI{1363}{\per\cm}) optical phonon mode at the center of the Brillouin zone. \cite{Serrano2007} They are also characterized by a monotonous decay in intensity upon increasing $n$. We thus attribute these new lines to three phonon replica of the dominant $S$ exciton recombinations. We notice that the energy of the second phonon replica almost  coincides with the $D4$ line. As a consequence, the $D/S$ ratio as defined here, has an intrinsic lower value, which corresponds to the ratio between the $n=2$ replica and the fundamental $S$ line. The observation of phonon replica attests to the existence of a strong electron-phonon coupling occurring in this material, as evidenced previously on a deep impurity band. \cite{Silly2007}

\subsection{Thickness of the exfoliated layers}

\begin{figure*}
	\centering
	\subfloat{
		\includegraphics[scale=0.45]{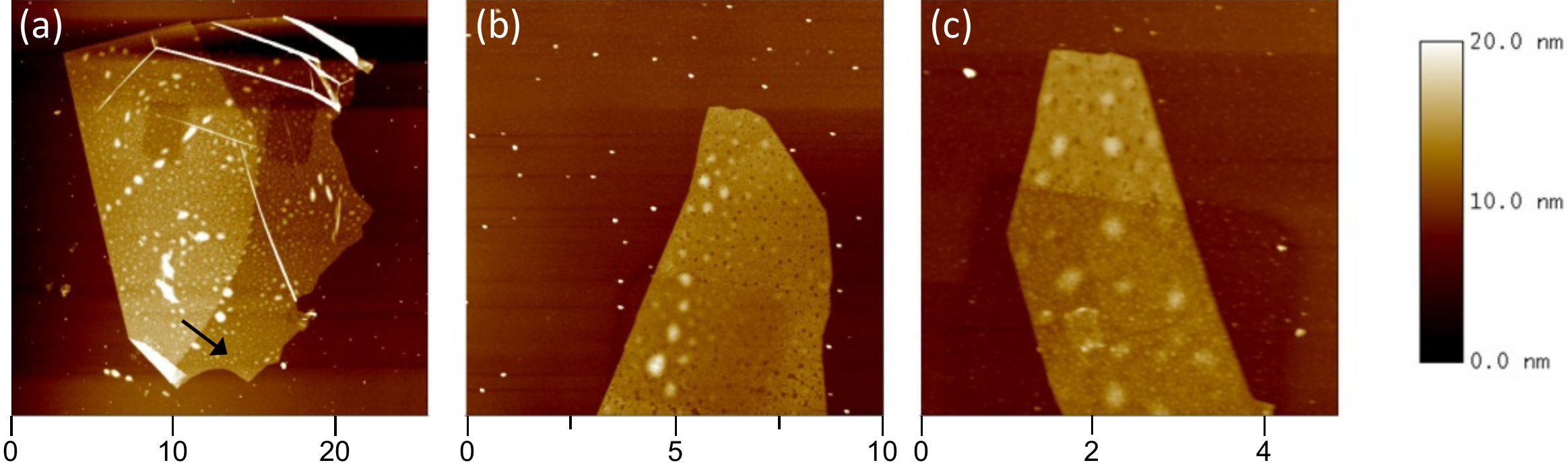}\label{AFM_A}}
	\subfloat{
		\label{AFM_B}}
	\subfloat{
		\label{AFM_C}}
	\subfloat{
		\includegraphics[scale=0.5]{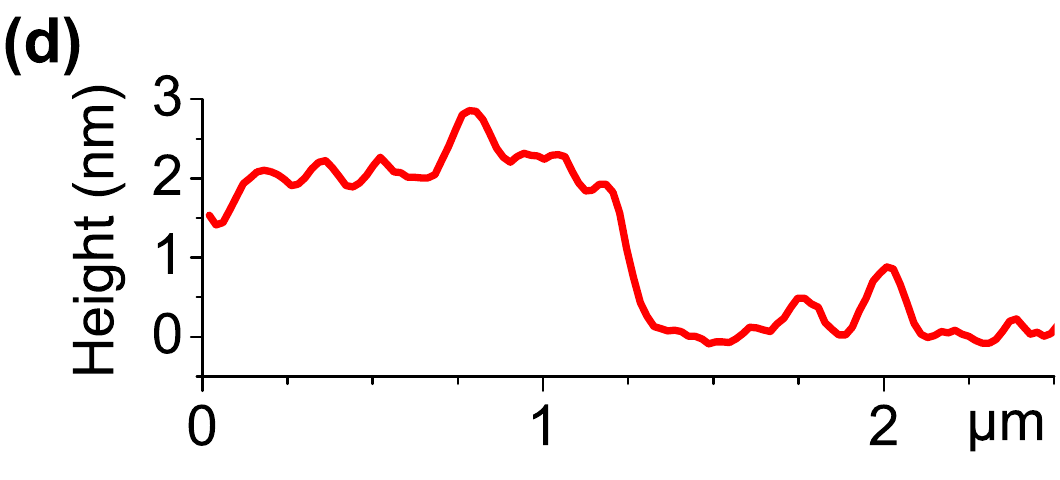}\label{profile}}
	\caption{AFM images of exfoliated $h$-BN layers transferred onto a \ce{SiO2}/Si substrate: (a) sample A; (b) sample B; (c) sample C. The image size unit is the micrometer. (d) Profile taken across the fold back part of the sheet A, along the arrow shown in (a).}
	\label{AFM}
\end{figure*}

Turning now to the  sheets exfoliated from the bulk material studied in Section IIIA, several sheets have been studied from which we show three representative samples in  \Cref{AFM}. The surface of each $h$-BN flake was scanned by AFM in the tapping mode in order to measure its thickness, consecutively to the CL studies. The sheet surface is not completely flat, with some prominent dots being visible. Most probably they are adhesive tape glue residues from the exfoliation process. Some small 10-\SI{100}{nm} size holes are also observable. It is not clear whether they were already in the bulk material or whether they are due to electron exposure as reported in the literature. Indeed when etching under an electron beam in a transmission electron microscope, even at a low acceleration voltage, they are  often observed. \cite{Jin2009,Suenaga12,Alem12,Pan12} Aside from this, the flakes appear to be well spread on top of the \ce{SiO2} substrate.

\begin{table*}
	\centering
	\begin{tabular}{|l|c|c|c|}
	\hline
	 & Sample A & Sample B & Sample C  \\ \hline
	Average thickness (nm)     & $2.1 \pm 0.1$ (5) & $2.8 \pm 0.4$ (5) & $3.4 \pm 0.27$ (4) \\ \hline
	Number of atomic layers    & 6 & 8 & 10  \\ \hline
	\end{tabular}
	\caption{Summary of AFM thickness measurements on exfoliated $h$-BN layers shown in \Cref{AFM}. The number of measurements over which the average and the standard deviation are calculated is indicated in parenthesis.}
	\label{table}
\end{table*}

Particular attention has been paid to the thickness determination because of the water adsorbed at the \ce{SiO2} surface and trapped under the BN flakes. This problem is well known in the case of graphene on \ce{SiO2} \cite{Ishigami2007,Nagashio11} and already mentioned for $h$-BN on \ce{SiO2}. \cite{Gorbachev11} Due to the presence of the water layer, the height profiles recorded by AFM on steps at the flake edges overestimate the $h$-BN thickness. The special geometry of sample A, which displays a folded part on itself, gives a key to solving this problem. The step height of the folded part provides a thickness measurement independent of the amount of water trapped between $h$-BN and  \ce{SiO2}, and this measurement has been used as the reference for the thickness measurement as follows. The $h$-BN flake thickness is found to be equal to a value of \SI{2.1}{nm} averaged over five  measurements of the step height along the edge (see  \Cref{profile}). The standard deviation is found to be below  \SI{0.3305}{nm}, the interplanar distance along the $c$ axis in $h$-BN, indicating that the thickness of flake A is homogeneous. Knowing the $h$-BN interplanar distance and assuming an integer value of the layer number, we consider that the exfoliated sample A is  6 atomic layers thick. Then the water layer thickness is deduced by difference with the step edges and is found equal to  $\SI{1.9}{nm} \pm \SI{0.25}{nm}$ over seven measurements. The thickness of samples B (\SI{8}{ML}) and C (\SI{10}{ML}) is measured assuming a homogenous water layer thickness over the wafer. The results are summarized in \Cref{table}.

One can notice that the areas scanned during the CL acquisition appear as dark rectangles in AFM images, which indicates a lower height. This could be interpreted by an electron beam induced etching. As we can also observe a lower height on the \ce{SiO2}, it is more probably a reduction of the volume of the amorphous silica under irradiation. \cite{Kalceff1996}


\subsection{Excitonic recombinations in $h$-BN exfoliated layers}

\begin{figure}
	\centering
	\includegraphics[scale=0.9]{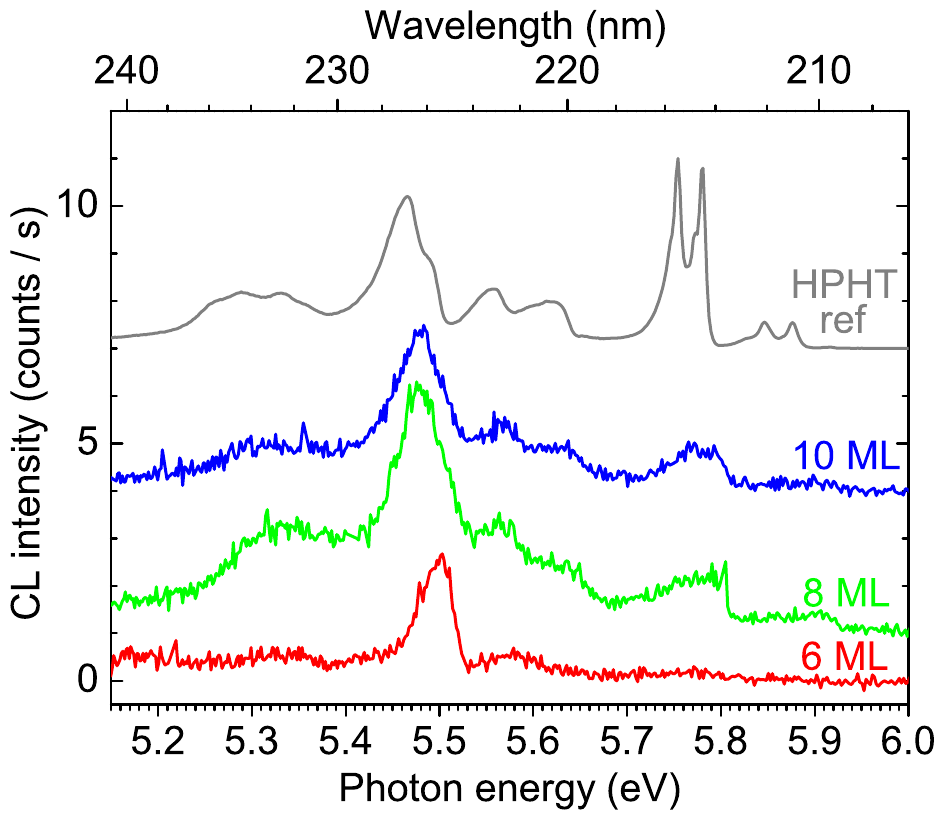}
	\caption{Near-band-edge exciton recombinations from exfoliated $h$-BN layer A (\SI{6}{ML}), B (\SI{8}{ML}) and C (\SI{10}{ML}) compared to the normalized reference spectrum (from the HPHT crystal). The zero of the CL spectra was shifted for clarity.}
	\label{flake}
\end{figure}

The CL spectra of the three exfoliated $h$-BN layers presented above are shown in  \Cref{flake}. First, we observe that the CL intensity decreases when decreasing the $h$-BN layer thickness. Since a  nanometer-thick $h$-BN layer is transparent to the electron beam, the electron-hole generation in such a thin layer is directly proportional to its thickness. In spite of this effect, exfoliated layers show a significant luminescence in comparison to their nanometric thickness, which is a result in itself. Indeed in semiconductor nanostructures, the surface to volume ratio, being extremely large, often results in non radiative surface recombinations, responsible for the quenching of luminescence. \cite{Demichel10,Calarco11}  Our results indicate that such a surface effect is weak in $h$-BN,  probably due to the $sp^2$ character of the bonding combined with the strong localization of the exciton at the atomic layer scale. This is an advantage of this material compared to other semiconductors, for which passivating the surface is necessary for enhancing the luminescence. \cite{Hines1996,Peng1997,Couto12,Titova2006}

The CL spectra of exfoliated layers are dominated by the $D4$ and $D2$ lines related to the $D$ series. Weaker and unresolved $S3-S4$ and $S1-S2$ lines correspond to the $S$ series, clearly attributed by comparison with the reference HPHT spectrum. Looking in more detail at the excitonic recombination emission, we observe a shift from  \SI{5.471}{eV} for the reference spectrum to  5.479, 5.477 and \SI{5.499}{eV} for exfoliated samples C (\SI{10}{ML}), B (\SI{8}{ML}) and A (\SI{6}{ML}) respectively. For the 8 and \SI{10}{ML}-thick samples, the blueshift is observed for the defect-related $D$ lines but also for the $S$ lines, indicating a modification in the intrinsic exciton emission energy (change in bandgap and/or in exciton binding energy). For the thinnest sample, the intrinsic $S$ line emission is not observed possibly because of a too low signal-to-noise ratio. 

To interpret our results, it has  to be emphasized first that a calculation including excitonic effects still needs to be performed to describe the influence of the number of atomic layers on $h$-BN optical properties. However, inside bulk $h$-BN crystals, the strong localization of the exciton wavefunction around a single atomic basal plane has been evidenced by Arnaud \textit{et al.} \cite{Arnaud2006}  It indicates that the regime of quantum confinement is still far to be reached with BN sheets in the range of 6-\SI{10}{ML}, and that exciton recombination properties are probably close to the ones of bulk $h$-BN.

It is also interesting to consider the theoretical work of Wirtz \textit{et al.} \cite{Wirtz2006}  It deals with the optical response of a single sheet of $h$-BN as a function of the intersheet distance. When the interdistance between $h$-BN planes becomes large, the situation tends to uncoupled BN sheets of atomic thickness.  The calculation results as follows: The exciton binding energy increases when the intersheet distance increases, but  simultaneously the quasiparticle band gap increases, so that the net result is a weak blueshift, about \SI{200}{meV}, in the error bar of such calculations. Similar effects have  been predicted in the case of BN nanotubes when their diameter decreases, \cite{Wirtz2006,ParkBN2006} and more recently in nanoribbons. \cite{Wang11} These arguments indicate that the change of the exciton recombination energy in monoatomic BN sheets should be weak compared to the bulk and should not exceed a few tenths of eV anyway.

Our experimental results show a slight increase  ($\approx$ \SI{30}{meV}) in the $D$-series exciton recombination energy from the bulk to the thinnest sample (\SI{6}{ML}), consistent with these theoretical predictions. Further calculations as well as experiments for less than six layers would be of great interest in the future to see if larger effects are observed.

\begin{figure}
	\centering
	\includegraphics[scale=0.9]{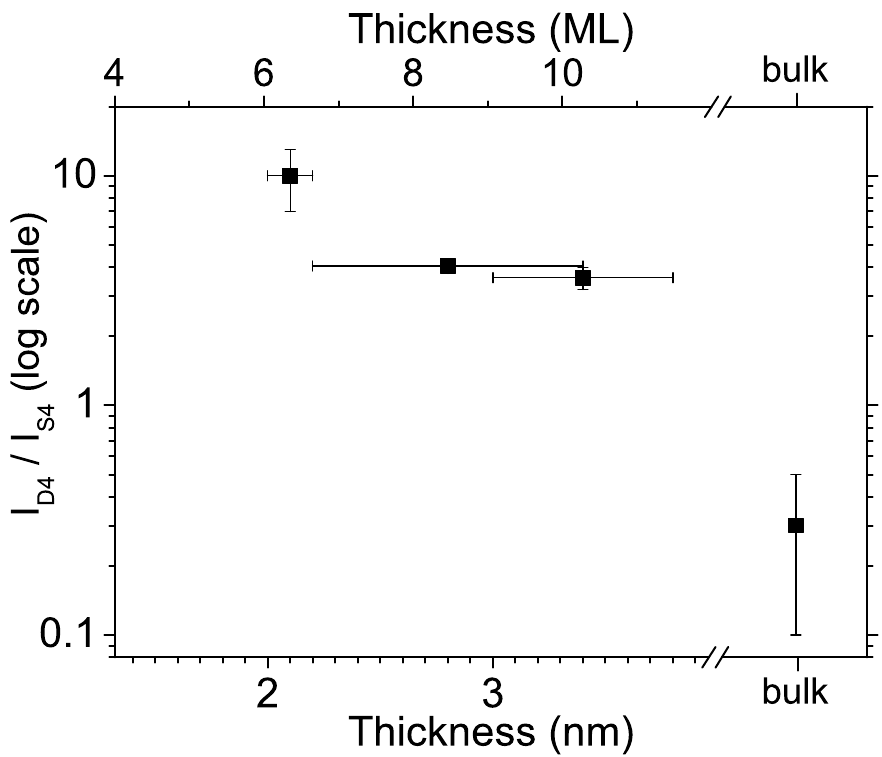}
	\caption{$D/S$ ratio in the three exfoliated layers, compared with the bulk material, consisting of the crystallites from which the layers were exfoliated.}
	\label{DSratio}
\end{figure}

When comparing the exfoliated materials with the bulk from which they were exfoliated, it is remarkable that excitonic recombinations of exfoliated layers are systematically dominated by the $D$ series. In the  bulk source materials, various crystallites have been investigated. We found that the $D/S$ ratio varies from one crystallite to another but stays between 0.2 and 0.5. By contrast, the $D/S$ ratio is found to be equal to about 4 in the 8 and \SI{10}{ML} sheets (samples B and C) and to about 10 in the  \SI{6}{ML} sheet (sample A), that is hundred times larger than in the bulk crystallites. These results are summarized in \Cref{DSratio}. As discussed previously, this increase of the $D/S$ ratio indicates a dramatic increase in the density of defects. They are probably induced by the exfoliation process, even if we cannot exclude layer deformation induced by the water layer freezing arising during CL experiments at \SI{20}{K}, or etching under the electron beam. These results raise the question of the suitability of the mechanical cleavage exfoliation process. $h$-BN has indeed been proven to be extremely fragile and subject, for instance, to easy stacking fault generation.\cite{WatanabeAPL2006} The exfoliation process may be a violent mechanical experiment for such a material and can be obviously incriminated as the source of new defects responsible for the $D/S$ increase. The nature of the substrate is also questionable. The roughness and surface charges of the used \ce{SiO2} substrate can affect the $h$-BN band structure and degrade some properties as for graphene. \cite{Gannett11}

\section{Conclusion}

In conclusion, we have studied high quality $h$-BN crystallites by cathodoluminescence and compared them with exfoliated sheets in the near-band edge region around \SI{5.5}{eV}. First owing to the high quality of the crystallites, we could clearly discriminate the luminescence of high structural quality areas (dominated by the $S$ series) and that of defective zones (dominated by the $D$ series). From this, for  evaluating the structural defect density in $h$-BN samples, we define a simple quantitative parameter as the ratio of the intensity of the $D$ series over that of the $S$ series. Furthermore the observation of high quality areas has revealed new intrinsic features, attributed to phonon replica of the excitonic recombinations.

Second, investigation of the emission of exfoliated sheets has shown that the reduction in thickness induces a slight increase of the exciton emission energy. These observations are consistent with theoretical results obtained on single layers  and on BN nanotubes. This explanation still deserves to be confirmed as one cannot exclude for the moment surface and strain effects. The dominant defect-related emission of exfoliated samples also questions the exfoliation process and the role played by the substrate. This indicates that  special care should be taken in the sample preparation. Presently, our measurements evidence that in $h$-BN, exfoliation can degrade some of the properties and should be used as a guideline to further optimize this process. Further experiments on free-standing sheets are thus in progress, with the advantage to make possible correlations between the emission and structural features owing to TEM observations. More generally this work reports luminescence spectra of a semiconductor only a few atomic layers thick, which is not common and can be of great interest for future optical applications.

\acknowledgements

T. Taniguchi and K. Watanabe from NIMS, Japan, are warmly acknowledged for providing one of their HPHT crystals.
The authors would like to thank S. Pouget from CEA Grenoble for x-ray experiments, C.Vilar from GEMaC for her technical help on the cathodoluminescence-SEM set-up, and H. Mariette from the NPSC for fruitful discussions. The research leading to these results has received funding from the European Union Seventh Framework Programme under grant agreement n\textsuperscript{o}604391 Graphene Flagship. It  has been also supported by grant from the Mission Interdisciplinaire of the CNRS (Challenge ``Graphene'' of the program G3N) and by the Federative Research Program ``Graphene'' of ONERA. A.P. thanks C'Nano Rh\^one-Alpes and Ile-de-France for financial support.


\begin{thebibliography}{0}%
\makeatletter
\providecommand \@ifxundefined [1]{%
 \@ifx{#1\undefined}
}%
\providecommand \@ifnum [1]{%
 \ifnum #1\expandafter \@firstoftwo
 \else \expandafter \@secondoftwo
 \fi
}%
\providecommand \@ifx [1]{%
 \ifx #1\expandafter \@firstoftwo
 \else \expandafter \@secondoftwo
 \fi
}%
\providecommand \natexlab [1]{#1}%
\providecommand \enquote  [1]{``#1''}%
\providecommand \bibnamefont  [1]{#1}%
\providecommand \bibfnamefont [1]{#1}%
\providecommand \citenamefont [1]{#1}%
\providecommand \href@noop [0]{\@secondoftwo}%
\providecommand \href [0]{\begingroup \@sanitize@url \@href}%
\providecommand \@href[1]{\@@startlink{#1}\@@href}%
\providecommand \@@href[1]{\endgroup#1\@@endlink}%
\providecommand \@sanitize@url [0]{\catcode `\\12\catcode `\$12\catcode
  `\&12\catcode `\#12\catcode `\^12\catcode `\_12\catcode `\%12\relax}%
\providecommand \@@startlink[1]{}%
\providecommand \@@endlink[0]{}%
\providecommand \url  [0]{\begingroup\@sanitize@url \@url }%
\providecommand \@url [1]{\endgroup\@href {#1}{\urlprefix }}%
\providecommand \urlprefix  [0]{URL }%
\providecommand \Eprint [0]{\href }%
\providecommand \doibase [0]{http://dx.doi.org/}%
\providecommand \selectlanguage [0]{\@gobble}%
\providecommand \bibinfo  [0]{\@secondoftwo}%
\providecommand \bibfield  [0]{\@secondoftwo}%
\providecommand \translation [1]{[#1]}%
\providecommand \BibitemOpen [0]{}%
\providecommand \bibitemStop [0]{}%
\providecommand \bibitemNoStop [0]{.\EOS\space}%
\providecommand \EOS [0]{\spacefactor3000\relax}%
\providecommand \BibitemShut  [1]{\csname bibitem#1\endcsname}%
\let\auto@bib@innerbib\@empty
\end{thebibliography}%


\begin{thebibliography}{59}%
\makeatletter
\providecommand \@ifxundefined [1]{%
 \@ifx{#1\undefined}
}%
\providecommand \@ifnum [1]{%
 \ifnum #1\expandafter \@firstoftwo
 \else \expandafter \@secondoftwo
 \fi
}%
\providecommand \@ifx [1]{%
 \ifx #1\expandafter \@firstoftwo
 \else \expandafter \@secondoftwo
 \fi
}%
\providecommand \natexlab [1]{#1}%
\providecommand \enquote  [1]{``#1''}%
\providecommand \bibnamefont  [1]{#1}%
\providecommand \bibfnamefont [1]{#1}%
\providecommand \citenamefont [1]{#1}%
\providecommand \href@noop [0]{\@secondoftwo}%
\providecommand \href [0]{\begingroup \@sanitize@url \@href}%
\providecommand \@href[1]{\@@startlink{#1}\@@href}%
\providecommand \@@href[1]{\endgroup#1\@@endlink}%
\providecommand \@sanitize@url [0]{\catcode `\\12\catcode `\$12\catcode
  `\&12\catcode `\#12\catcode `\^12\catcode `\_12\catcode `\%12\relax}%
\providecommand \@@startlink[1]{}%
\providecommand \@@endlink[0]{}%
\providecommand \url  [0]{\begingroup\@sanitize@url \@url }%
\providecommand \@url [1]{\endgroup\@href {#1}{\urlprefix }}%
\providecommand \urlprefix  [0]{URL }%
\providecommand \Eprint [0]{\href }%
\providecommand \doibase [0]{http://dx.doi.org/}%
\providecommand \selectlanguage [0]{\@gobble}%
\providecommand \bibinfo  [0]{\@secondoftwo}%
\providecommand \bibfield  [0]{\@secondoftwo}%
\providecommand \translation [1]{[#1]}%
\providecommand \BibitemOpen [0]{}%
\providecommand \bibitemStop [0]{}%
\providecommand \bibitemNoStop [0]{.\EOS\space}%
\providecommand \EOS [0]{\spacefactor3000\relax}%
\providecommand \BibitemShut  [1]{\csname bibitem#1\endcsname}%
\let\auto@bib@innerbib\@empty
\bibitem [{\citenamefont {Chopra}\ \emph {et~al.}(1995)\citenamefont {Chopra},
  \citenamefont {Luyken}, \citenamefont {Cherrey}, \citenamefont {Crespi},
  \citenamefont {Cohen}, \citenamefont {Louie},\ and\ \citenamefont
  {Zettl}}]{Chopra1995}%
  \BibitemOpen
  \bibfield  {author} {\bibinfo {author} {\bibfnamefont {N.~G.}\ \bibnamefont
  {Chopra}}, \bibinfo {author} {\bibfnamefont {R.~J.}\ \bibnamefont {Luyken}},
  \bibinfo {author} {\bibfnamefont {K.}~\bibnamefont {Cherrey}}, \bibinfo
  {author} {\bibfnamefont {V.~H.}\ \bibnamefont {Crespi}}, \bibinfo {author}
  {\bibfnamefont {M.~L.}\ \bibnamefont {Cohen}}, \bibinfo {author}
  {\bibfnamefont {S.~G.}\ \bibnamefont {Louie}}, \ and\ \bibinfo {author}
  {\bibfnamefont {A.}~\bibnamefont {Zettl}},\ }\href@noop {} {\bibfield
  {journal} {\bibinfo  {journal} {Science}\ }\textbf {\bibinfo {volume}
  {269}},\ \bibinfo {pages} {966} (\bibinfo {year} {1995})}\BibitemShut
  {NoStop}%
\bibitem [{\citenamefont {Loiseau}\ \emph {et~al.}(1996)\citenamefont
  {Loiseau}, \citenamefont {Willaime}, \citenamefont {Demoncy}, \citenamefont
  {Hug},\ and\ \citenamefont {Pascard}}]{Loiseau1996}%
  \BibitemOpen
  \bibfield  {author} {\bibinfo {author} {\bibfnamefont {A.}~\bibnamefont
  {Loiseau}}, \bibinfo {author} {\bibfnamefont {F.}~\bibnamefont {Willaime}},
  \bibinfo {author} {\bibfnamefont {N.}~\bibnamefont {Demoncy}}, \bibinfo
  {author} {\bibfnamefont {G.}~\bibnamefont {Hug}}, \ and\ \bibinfo {author}
  {\bibfnamefont {H.}~\bibnamefont {Pascard}},\ }\href@noop {} {\bibfield
  {journal} {\bibinfo  {journal} {Phys. Rev. Lett.}\ }\textbf {\bibinfo
  {volume} {76}},\ \bibinfo {pages} {4737} (\bibinfo {year}
  {1996})}\BibitemShut {NoStop}%
\bibitem [{\citenamefont {Shi}\ \emph {et~al.}(2010)\citenamefont {Shi},
  \citenamefont {Hamsen}, \citenamefont {Jia}, \citenamefont {Kim},
  \citenamefont {Reina}, \citenamefont {Hofmann}, \citenamefont {Hsu},
  \citenamefont {Zhang}, \citenamefont {Li}, \citenamefont {Juang},
  \citenamefont {Dresselhaus}, \citenamefont {Li},\ and\ \citenamefont
  {Kong}}]{Shi10}%
  \BibitemOpen
  \bibfield  {author} {\bibinfo {author} {\bibfnamefont {Y.}~\bibnamefont
  {Shi}}, \bibinfo {author} {\bibfnamefont {C.}~\bibnamefont {Hamsen}},
  \bibinfo {author} {\bibfnamefont {X.}~\bibnamefont {Jia}}, \bibinfo {author}
  {\bibfnamefont {K.~K.}\ \bibnamefont {Kim}}, \bibinfo {author} {\bibfnamefont
  {A.}~\bibnamefont {Reina}}, \bibinfo {author} {\bibfnamefont
  {M.}~\bibnamefont {Hofmann}}, \bibinfo {author} {\bibfnamefont {A.~L.}\
  \bibnamefont {Hsu}}, \bibinfo {author} {\bibfnamefont {K.}~\bibnamefont
  {Zhang}}, \bibinfo {author} {\bibfnamefont {H.}~\bibnamefont {Li}}, \bibinfo
  {author} {\bibfnamefont {Z.-Y.}\ \bibnamefont {Juang}}, \bibinfo {author}
  {\bibfnamefont {M.~S.}\ \bibnamefont {Dresselhaus}}, \bibinfo {author}
  {\bibfnamefont {L.-J.}\ \bibnamefont {Li}}, \ and\ \bibinfo {author}
  {\bibfnamefont {J.}~\bibnamefont {Kong}},\ }\href {\doibase
  10.1021/nl1023707} {\bibfield  {journal} {\bibinfo  {journal} {Nano Lett.}\
  }\textbf {\bibinfo {volume} {10}},\ \bibinfo {pages} {4134} (\bibinfo {year}
  {2010})}\BibitemShut {NoStop}%
\bibitem [{\citenamefont {Song}\ \emph {et~al.}(2010)\citenamefont {Song},
  \citenamefont {Ci}, \citenamefont {Lu}, \citenamefont {Sorokin},
  \citenamefont {Jin}, \citenamefont {Ni}, \citenamefont {Kvashnin},
  \citenamefont {Kvashnin}, \citenamefont {Lou}, \citenamefont {Yakobson},\
  and\ \citenamefont {Ajayan}}]{Song10}%
  \BibitemOpen
  \bibfield  {author} {\bibinfo {author} {\bibfnamefont {L.}~\bibnamefont
  {Song}}, \bibinfo {author} {\bibfnamefont {L.}~\bibnamefont {Ci}}, \bibinfo
  {author} {\bibfnamefont {H.}~\bibnamefont {Lu}}, \bibinfo {author}
  {\bibfnamefont {P.}~\bibnamefont {Sorokin}}, \bibinfo {author} {\bibfnamefont
  {C.}~\bibnamefont {Jin}}, \bibinfo {author} {\bibfnamefont {J.}~\bibnamefont
  {Ni}}, \bibinfo {author} {\bibfnamefont {A.}~\bibnamefont {Kvashnin}},
  \bibinfo {author} {\bibfnamefont {D.}~\bibnamefont {Kvashnin}}, \bibinfo
  {author} {\bibfnamefont {J.}~\bibnamefont {Lou}}, \bibinfo {author}
  {\bibfnamefont {B.}~\bibnamefont {Yakobson}}, \ and\ \bibinfo {author}
  {\bibfnamefont {P.~M.}\ \bibnamefont {Ajayan}},\ }\href {\doibase
  10.1021/nl1022139} {\bibfield  {journal} {\bibinfo  {journal} {Nano Lett.}\
  }\textbf {\bibinfo {volume} {10}},\ \bibinfo {pages} {3209} (\bibinfo {year}
  {2010})}\BibitemShut {NoStop}%
\bibitem [{\citenamefont {Burson}\ \emph {et~al.}(2013)\citenamefont {Burson},
  \citenamefont {Cullen}, \citenamefont {Adam}, \citenamefont {Dean},
  \citenamefont {Watanabe}, \citenamefont {Taniguchi}, \citenamefont {Kim},\
  and\ \citenamefont {Fuhrer}}]{Burson13}%
  \BibitemOpen
  \bibfield  {author} {\bibinfo {author} {\bibfnamefont {K.~M.}\ \bibnamefont
  {Burson}}, \bibinfo {author} {\bibfnamefont {W.~G.}\ \bibnamefont {Cullen}},
  \bibinfo {author} {\bibfnamefont {S.}~\bibnamefont {Adam}}, \bibinfo {author}
  {\bibfnamefont {C.~R.}\ \bibnamefont {Dean}}, \bibinfo {author}
  {\bibfnamefont {K.}~\bibnamefont {Watanabe}}, \bibinfo {author}
  {\bibfnamefont {T.}~\bibnamefont {Taniguchi}}, \bibinfo {author}
  {\bibfnamefont {P.}~\bibnamefont {Kim}}, \ and\ \bibinfo {author}
  {\bibfnamefont {M.~S.}\ \bibnamefont {Fuhrer}},\ }\href {\doibase
  10.1021/nl4012529} {\bibfield  {journal} {\bibinfo  {journal} {Nano Lett.}\
  }\textbf {\bibinfo {volume} {13}},\ \bibinfo {pages} {3576} (\bibinfo {year}
  {2013})}\BibitemShut {NoStop}%
\bibitem [{\citenamefont {Xue}\ \emph {et~al.}(2011)\citenamefont {Xue},
  \citenamefont {Sanchez-Yamagishi}, \citenamefont {Bulmash}, \citenamefont
  {Jacquod}, \citenamefont {Deshpande}, \citenamefont {Watanabe}, \citenamefont
  {Taniguchi}, \citenamefont {Jarillo-Herrero},\ and\ \citenamefont
  {LeRoy}}]{Xue11}%
  \BibitemOpen
  \bibfield  {author} {\bibinfo {author} {\bibfnamefont {J.}~\bibnamefont
  {Xue}}, \bibinfo {author} {\bibfnamefont {J.}~\bibnamefont
  {Sanchez-Yamagishi}}, \bibinfo {author} {\bibfnamefont {D.}~\bibnamefont
  {Bulmash}}, \bibinfo {author} {\bibfnamefont {P.}~\bibnamefont {Jacquod}},
  \bibinfo {author} {\bibfnamefont {A.}~\bibnamefont {Deshpande}}, \bibinfo
  {author} {\bibfnamefont {K.}~\bibnamefont {Watanabe}}, \bibinfo {author}
  {\bibfnamefont {T.}~\bibnamefont {Taniguchi}}, \bibinfo {author}
  {\bibfnamefont {P.}~\bibnamefont {Jarillo-Herrero}}, \ and\ \bibinfo {author}
  {\bibfnamefont {B.}~\bibnamefont {LeRoy}},\ }\href@noop {} {\bibfield
  {journal} {\bibinfo  {journal} {Nat. Mater.}\ }\textbf {\bibinfo {volume}
  {10}},\ \bibinfo {pages} {282} (\bibinfo {year} {2011})}\BibitemShut
  {NoStop}%
\bibitem [{\citenamefont {Decker}\ \emph {et~al.}(2011)\citenamefont {Decker},
  \citenamefont {Wang}, \citenamefont {Brar}, \citenamefont {Regan},
  \citenamefont {Tsai}, \citenamefont {Wu}, \citenamefont {Gannett},
  \citenamefont {Zettl},\ and\ \citenamefont {Crommie}}]{Decker11}%
  \BibitemOpen
  \bibfield  {author} {\bibinfo {author} {\bibfnamefont {R.}~\bibnamefont
  {Decker}}, \bibinfo {author} {\bibfnamefont {Y.}~\bibnamefont {Wang}},
  \bibinfo {author} {\bibfnamefont {V.~W.}\ \bibnamefont {Brar}}, \bibinfo
  {author} {\bibfnamefont {W.}~\bibnamefont {Regan}}, \bibinfo {author}
  {\bibfnamefont {H.-Z.}\ \bibnamefont {Tsai}}, \bibinfo {author}
  {\bibfnamefont {Q.}~\bibnamefont {Wu}}, \bibinfo {author} {\bibfnamefont
  {W.}~\bibnamefont {Gannett}}, \bibinfo {author} {\bibfnamefont
  {A.}~\bibnamefont {Zettl}}, \ and\ \bibinfo {author} {\bibfnamefont {M.~F.}\
  \bibnamefont {Crommie}},\ }\href {\doibase 10.1021/nl2005115} {\bibfield
  {journal} {\bibinfo  {journal} {Nano Lett.}\ }\textbf {\bibinfo {volume}
  {11}},\ \bibinfo {pages} {2291} (\bibinfo {year} {2011})}\BibitemShut
  {NoStop}%
\bibitem [{\citenamefont {Zunger}\ \emph {et~al.}(1976)\citenamefont {Zunger},
  \citenamefont {Katzir},\ and\ \citenamefont {Halperin}}]{Zunger1976}%
  \BibitemOpen
  \bibfield  {author} {\bibinfo {author} {\bibfnamefont {A.}~\bibnamefont
  {Zunger}}, \bibinfo {author} {\bibfnamefont {A.}~\bibnamefont {Katzir}}, \
  and\ \bibinfo {author} {\bibfnamefont {A.}~\bibnamefont {Halperin}},\
  }\href@noop {} {\bibfield  {journal} {\bibinfo  {journal} {Phys. Rev. B}\
  }\textbf {\bibinfo {volume} {13}},\ \bibinfo {pages} {5560} (\bibinfo {year}
  {1976})}\BibitemShut {NoStop}%
\bibitem [{\citenamefont {Museur}\ \emph {et~al.}(2011)\citenamefont {Museur},
  \citenamefont {Brasse}, \citenamefont {Pierret}, \citenamefont {Maine},
  \citenamefont {Attal-Tretout}, \citenamefont {Ducastelle}, \citenamefont
  {Loiseau}, \citenamefont {Barjon}, \citenamefont {Watanabe}, \citenamefont
  {Taniguchi},\ and\ \citenamefont {Kanaev}}]{Museur11}%
  \BibitemOpen
  \bibfield  {author} {\bibinfo {author} {\bibfnamefont {L.}~\bibnamefont
  {Museur}}, \bibinfo {author} {\bibfnamefont {G.}~\bibnamefont {Brasse}},
  \bibinfo {author} {\bibfnamefont {A.}~\bibnamefont {Pierret}}, \bibinfo
  {author} {\bibfnamefont {S.}~\bibnamefont {Maine}}, \bibinfo {author}
  {\bibfnamefont {B.}~\bibnamefont {Attal-Tretout}}, \bibinfo {author}
  {\bibfnamefont {F.}~\bibnamefont {Ducastelle}}, \bibinfo {author}
  {\bibfnamefont {A.}~\bibnamefont {Loiseau}}, \bibinfo {author} {\bibfnamefont
  {J.}~\bibnamefont {Barjon}}, \bibinfo {author} {\bibfnamefont
  {K.}~\bibnamefont {Watanabe}}, \bibinfo {author} {\bibfnamefont
  {T.}~\bibnamefont {Taniguchi}}, \ and\ \bibinfo {author} {\bibfnamefont
  {A.}~\bibnamefont {Kanaev}},\ }\href {\doibase 10.1002/pssr.201105190}
  {\bibfield  {journal} {\bibinfo  {journal} {Phys. Status Solidi RRL}\
  }\textbf {\bibinfo {volume} {5}},\ \bibinfo {pages} {214} (\bibinfo {year}
  {2011})}\BibitemShut {NoStop}%
\bibitem [{\citenamefont {Blase}\ \emph {et~al.}(1995)\citenamefont {Blase},
  \citenamefont {Rubio}, \citenamefont {Louie},\ and\ \citenamefont
  {Cohen}}]{Blase1995}%
  \BibitemOpen
  \bibfield  {author} {\bibinfo {author} {\bibfnamefont {X.}~\bibnamefont
  {Blase}}, \bibinfo {author} {\bibfnamefont {A.}~\bibnamefont {Rubio}},
  \bibinfo {author} {\bibfnamefont {S.~G.}\ \bibnamefont {Louie}}, \ and\
  \bibinfo {author} {\bibfnamefont {M.~L.}\ \bibnamefont {Cohen}},\ }\href@noop
  {} {\bibfield  {journal} {\bibinfo  {journal} {Phys. Rev. B}\ }\textbf
  {\bibinfo {volume} {51}},\ \bibinfo {pages} {6868} (\bibinfo {year}
  {1995})}\BibitemShut {NoStop}%
\bibitem [{\citenamefont {Dean}\ \emph {et~al.}(2010)\citenamefont {Dean},
  \citenamefont {Young}, \citenamefont {Meric}, \citenamefont {Lee},
  \citenamefont {Wang}, \citenamefont {Sorgenfrei}, \citenamefont {Watanabe},
  \citenamefont {Taniguchi}, \citenamefont {Kim}, \citenamefont {Shepard},\
  and\ \citenamefont {Hone}}]{Dean10}%
  \BibitemOpen
  \bibfield  {author} {\bibinfo {author} {\bibfnamefont {C.}~\bibnamefont
  {Dean}}, \bibinfo {author} {\bibfnamefont {A.}~\bibnamefont {Young}},
  \bibinfo {author} {\bibfnamefont {I.}~\bibnamefont {Meric}}, \bibinfo
  {author} {\bibfnamefont {C.}~\bibnamefont {Lee}}, \bibinfo {author}
  {\bibfnamefont {L.}~\bibnamefont {Wang}}, \bibinfo {author} {\bibfnamefont
  {S.}~\bibnamefont {Sorgenfrei}}, \bibinfo {author} {\bibfnamefont
  {K.}~\bibnamefont {Watanabe}}, \bibinfo {author} {\bibfnamefont
  {T.}~\bibnamefont {Taniguchi}}, \bibinfo {author} {\bibfnamefont
  {P.}~\bibnamefont {Kim}}, \bibinfo {author} {\bibfnamefont {K.}~\bibnamefont
  {Shepard}}, \ and\ \bibinfo {author} {\bibfnamefont {J.}~\bibnamefont
  {Hone}},\ }\href@noop {} {\bibfield  {journal} {\bibinfo  {journal} {Nat.
  Nanotechnol.}\ }\textbf {\bibinfo {volume} {5}},\ \bibinfo {pages} {722}
  (\bibinfo {year} {2010})}\BibitemShut {NoStop}%
\bibitem [{\citenamefont {Gannett}\ \emph {et~al.}(2011)\citenamefont
  {Gannett}, \citenamefont {Regan}, \citenamefont {Watanabe}, \citenamefont
  {Taniguchi}, \citenamefont {Crommie},\ and\ \citenamefont
  {Zettl}}]{Gannett11}%
  \BibitemOpen
  \bibfield  {author} {\bibinfo {author} {\bibfnamefont {W.}~\bibnamefont
  {Gannett}}, \bibinfo {author} {\bibfnamefont {W.}~\bibnamefont {Regan}},
  \bibinfo {author} {\bibfnamefont {K.}~\bibnamefont {Watanabe}}, \bibinfo
  {author} {\bibfnamefont {T.}~\bibnamefont {Taniguchi}}, \bibinfo {author}
  {\bibfnamefont {M.~F.}\ \bibnamefont {Crommie}}, \ and\ \bibinfo {author}
  {\bibfnamefont {A.}~\bibnamefont {Zettl}},\ }\href {\doibase
  10.1063/1.3599708} {\bibfield  {journal} {\bibinfo  {journal} {Appl. Phys.
  Lett.}\ }\textbf {\bibinfo {volume} {98}},\ \bibinfo {pages} {242105}
  (\bibinfo {year} {2011})}\BibitemShut {NoStop}%
\bibitem [{\citenamefont {Mayorov}\ \emph {et~al.}(2011)\citenamefont
  {Mayorov}, \citenamefont {Gorbachev}, \citenamefont {Morozov}, \citenamefont
  {Britnell}, \citenamefont {Jalil}, \citenamefont {Ponomarenko}, \citenamefont
  {Blake}, \citenamefont {Novoselov}, \citenamefont {Watanabe}, \citenamefont
  {Taniguchi},\ and\ \citenamefont {Geim}}]{Mayorov11}%
  \BibitemOpen
  \bibfield  {author} {\bibinfo {author} {\bibfnamefont {A.~S.}\ \bibnamefont
  {Mayorov}}, \bibinfo {author} {\bibfnamefont {R.~V.}\ \bibnamefont
  {Gorbachev}}, \bibinfo {author} {\bibfnamefont {S.~V.}\ \bibnamefont
  {Morozov}}, \bibinfo {author} {\bibfnamefont {L.}~\bibnamefont {Britnell}},
  \bibinfo {author} {\bibfnamefont {R.}~\bibnamefont {Jalil}}, \bibinfo
  {author} {\bibfnamefont {L.~A.}\ \bibnamefont {Ponomarenko}}, \bibinfo
  {author} {\bibfnamefont {P.}~\bibnamefont {Blake}}, \bibinfo {author}
  {\bibfnamefont {K.~S.}\ \bibnamefont {Novoselov}}, \bibinfo {author}
  {\bibfnamefont {K.}~\bibnamefont {Watanabe}}, \bibinfo {author}
  {\bibfnamefont {T.}~\bibnamefont {Taniguchi}}, \ and\ \bibinfo {author}
  {\bibfnamefont {A.~K.}\ \bibnamefont {Geim}},\ }\href {\doibase
  10.1021/nl200758b} {\bibfield  {journal} {\bibinfo  {journal} {Nano Lett.}\
  }\textbf {\bibinfo {volume} {11}},\ \bibinfo {pages} {2396} (\bibinfo {year}
  {2011})}\BibitemShut {NoStop}%
\bibitem [{\citenamefont {Zomer}\ \emph {et~al.}(2011)\citenamefont {Zomer},
  \citenamefont {Dash}, \citenamefont {Tombros},\ and\ \citenamefont {van
  Wees}}]{Zomer11}%
  \BibitemOpen
  \bibfield  {author} {\bibinfo {author} {\bibfnamefont {P.~J.}\ \bibnamefont
  {Zomer}}, \bibinfo {author} {\bibfnamefont {S.~P.}\ \bibnamefont {Dash}},
  \bibinfo {author} {\bibfnamefont {N.}~\bibnamefont {Tombros}}, \ and\
  \bibinfo {author} {\bibfnamefont {B.~J.}\ \bibnamefont {van Wees}},\ }\href
  {\doibase 10.1063/1.3665405} {\bibfield  {journal} {\bibinfo  {journal}
  {Appl. Phys. Lett.}\ }\textbf {\bibinfo {volume} {99}},\ \bibinfo {eid}
  {232104} (\bibinfo {year} {2011})}\BibitemShut {NoStop}%
\bibitem [{\citenamefont {Bolotin}\ \emph {et~al.}(2008)\citenamefont
  {Bolotin}, \citenamefont {Sikes}, \citenamefont {Hone}, \citenamefont
  {Stormer},\ and\ \citenamefont {Kim}}]{Bolotin2008}%
  \BibitemOpen
  \bibfield  {author} {\bibinfo {author} {\bibfnamefont {K.~I.}\ \bibnamefont
  {Bolotin}}, \bibinfo {author} {\bibfnamefont {K.~J.}\ \bibnamefont {Sikes}},
  \bibinfo {author} {\bibfnamefont {J.}~\bibnamefont {Hone}}, \bibinfo {author}
  {\bibfnamefont {H.~L.}\ \bibnamefont {Stormer}}, \ and\ \bibinfo {author}
  {\bibfnamefont {P.}~\bibnamefont {Kim}},\ }\href {\doibase
  10.1103/PhysRevLett.101.096802} {\bibfield  {journal} {\bibinfo  {journal}
  {Phys. Rev. Lett.}\ }\textbf {\bibinfo {volume} {101}},\ \bibinfo {pages}
  {096802} (\bibinfo {year} {2008})}\BibitemShut {NoStop}%
\bibitem [{\citenamefont {Du}\ \emph {et~al.}(2008)\citenamefont {Du},
  \citenamefont {Skachko}, \citenamefont {Barker},\ and\ \citenamefont
  {Andrei}}]{Du2008}%
  \BibitemOpen
  \bibfield  {author} {\bibinfo {author} {\bibfnamefont {X.}~\bibnamefont
  {Du}}, \bibinfo {author} {\bibfnamefont {I.}~\bibnamefont {Skachko}},
  \bibinfo {author} {\bibfnamefont {A.}~\bibnamefont {Barker}}, \ and\ \bibinfo
  {author} {\bibfnamefont {E.~Y.}\ \bibnamefont {Andrei}},\ }\href
  {http://dx.doi.org/10.1038/nnano.2008.199} {\bibfield  {journal} {\bibinfo
  {journal} {Nat Nano}\ }\textbf {\bibinfo {volume} {3}},\ \bibinfo {pages}
  {491} (\bibinfo {year} {2008})}\BibitemShut {NoStop}%
\bibitem [{\citenamefont {Britnell}\ \emph {et~al.}(2012)\citenamefont
  {Britnell}, \citenamefont {Gorbachev}, \citenamefont {Jalil}, \citenamefont
  {Belle}, \citenamefont {Schedin}, \citenamefont {Mishchenko}, \citenamefont
  {Georgiou}, \citenamefont {Katsnelson}, \citenamefont {Eaves}, \citenamefont
  {Morozov}, \citenamefont {Peres}, \citenamefont {Leist}, \citenamefont
  {Geim}, \citenamefont {Novoselov},\ and\ \citenamefont
  {Ponomarenko}}]{Britnell12}%
  \BibitemOpen
  \bibfield  {author} {\bibinfo {author} {\bibfnamefont {L.}~\bibnamefont
  {Britnell}}, \bibinfo {author} {\bibfnamefont {R.~V.}\ \bibnamefont
  {Gorbachev}}, \bibinfo {author} {\bibfnamefont {R.}~\bibnamefont {Jalil}},
  \bibinfo {author} {\bibfnamefont {B.~D.}\ \bibnamefont {Belle}}, \bibinfo
  {author} {\bibfnamefont {F.}~\bibnamefont {Schedin}}, \bibinfo {author}
  {\bibfnamefont {A.}~\bibnamefont {Mishchenko}}, \bibinfo {author}
  {\bibfnamefont {T.}~\bibnamefont {Georgiou}}, \bibinfo {author}
  {\bibfnamefont {M.~I.}\ \bibnamefont {Katsnelson}}, \bibinfo {author}
  {\bibfnamefont {L.}~\bibnamefont {Eaves}}, \bibinfo {author} {\bibfnamefont
  {S.~V.}\ \bibnamefont {Morozov}}, \bibinfo {author} {\bibfnamefont
  {N.~M.~R.}\ \bibnamefont {Peres}}, \bibinfo {author} {\bibfnamefont
  {J.}~\bibnamefont {Leist}}, \bibinfo {author} {\bibfnamefont {A.~K.}\
  \bibnamefont {Geim}}, \bibinfo {author} {\bibfnamefont {K.~S.}\ \bibnamefont
  {Novoselov}}, \ and\ \bibinfo {author} {\bibfnamefont {L.~A.}\ \bibnamefont
  {Ponomarenko}},\ }\href {\doibase 10.1126/science.1218461} {\bibfield
  {journal} {\bibinfo  {journal} {Science}\ }\textbf {\bibinfo {volume}
  {335}},\ \bibinfo {pages} {947} (\bibinfo {year} {2012})}\BibitemShut
  {NoStop}%
\bibitem [{\citenamefont {Ramasubramaniam}\ \emph {et~al.}(2011)\citenamefont
  {Ramasubramaniam}, \citenamefont {Naveh},\ and\ \citenamefont
  {Towe}}]{Ramasubramaniam11}%
  \BibitemOpen
  \bibfield  {author} {\bibinfo {author} {\bibfnamefont {A.}~\bibnamefont
  {Ramasubramaniam}}, \bibinfo {author} {\bibfnamefont {D.}~\bibnamefont
  {Naveh}}, \ and\ \bibinfo {author} {\bibfnamefont {E.}~\bibnamefont {Towe}},\
  }\href {\doibase 10.1021/nl1039499} {\bibfield  {journal} {\bibinfo
  {journal} {Nano Lett.}\ }\textbf {\bibinfo {volume} {11}},\ \bibinfo {pages}
  {1070} (\bibinfo {year} {2011})}\BibitemShut {NoStop}%
\bibitem [{\citenamefont {Jaffrennou}\ \emph {et~al.}(2007)\citenamefont
  {Jaffrennou}, \citenamefont {Barjon}, \citenamefont {Lauret}, \citenamefont
  {Attal-Tr{\'e}tout}, \citenamefont {Ducastelle},\ and\ \citenamefont
  {Loiseau}}]{JaffrennouJAP2007}%
  \BibitemOpen
  \bibfield  {author} {\bibinfo {author} {\bibfnamefont {P.}~\bibnamefont
  {Jaffrennou}}, \bibinfo {author} {\bibfnamefont {J.}~\bibnamefont {Barjon}},
  \bibinfo {author} {\bibfnamefont {J.-S.}\ \bibnamefont {Lauret}}, \bibinfo
  {author} {\bibfnamefont {B.}~\bibnamefont {Attal-Tr{\'e}tout}}, \bibinfo
  {author} {\bibfnamefont {F.}~\bibnamefont {Ducastelle}}, \ and\ \bibinfo
  {author} {\bibfnamefont {A.}~\bibnamefont {Loiseau}},\ }\href@noop {}
  {\bibfield  {journal} {\bibinfo  {journal} {J. Appl. Phys.}\ }\textbf
  {\bibinfo {volume} {102}},\ \bibinfo {pages} {116102} (\bibinfo {year}
  {2007})}\BibitemShut {NoStop}%
\bibitem [{\citenamefont {Watanabe}\ and\ \citenamefont
  {Taniguchi}(2009)}]{WatanabePRB2009}%
  \BibitemOpen
  \bibfield  {author} {\bibinfo {author} {\bibfnamefont {K.}~\bibnamefont
  {Watanabe}}\ and\ \bibinfo {author} {\bibfnamefont {T.}~\bibnamefont
  {Taniguchi}},\ }\href@noop {} {\bibfield  {journal} {\bibinfo  {journal}
  {Phys. Rev. B}\ }\textbf {\bibinfo {volume} {79}},\ \bibinfo {pages} {193104}
  (\bibinfo {year} {2009})}\BibitemShut {NoStop}%
\bibitem [{\citenamefont {Museur}\ and\ \citenamefont
  {Kanaev}(2008)}]{MuseurJAP2008}%
  \BibitemOpen
  \bibfield  {author} {\bibinfo {author} {\bibfnamefont {L.}~\bibnamefont
  {Museur}}\ and\ \bibinfo {author} {\bibfnamefont {A.}~\bibnamefont
  {Kanaev}},\ }\href@noop {} {\bibfield  {journal} {\bibinfo  {journal} {J.
  Appl. Phys.}\ }\textbf {\bibinfo {volume} {103}},\ \bibinfo {pages} {103520}
  (\bibinfo {year} {2008})}\BibitemShut {NoStop}%
\bibitem [{\citenamefont {Wirtz}\ \emph {et~al.}(2005)\citenamefont {Wirtz},
  \citenamefont {Marini}, \citenamefont {Gr{\"u}ning},\ and\ \citenamefont
  {Rubio}}]{WirtzArXiv2005}%
  \BibitemOpen
  \bibfield  {author} {\bibinfo {author} {\bibfnamefont {L.}~\bibnamefont
  {Wirtz}}, \bibinfo {author} {\bibfnamefont {A.}~\bibnamefont {Marini}},
  \bibinfo {author} {\bibfnamefont {M.}~\bibnamefont {Gr{\"u}ning}}, \ and\
  \bibinfo {author} {\bibfnamefont {A.}~\bibnamefont {Rubio}},\ }\href@noop {}
  {\bibfield  {journal} {\bibinfo  {journal} {ArXiv}\ }\textbf {\bibinfo
  {volume} {:cond-mat/0508421}} (\bibinfo {year} {2005})}\BibitemShut {NoStop}%
\bibitem [{\citenamefont {Arnaud}\ \emph {et~al.}(2006)\citenamefont {Arnaud},
  \citenamefont {Leb{\`e}gue}, \citenamefont {Rabiller},\ and\ \citenamefont
  {Alouani}}]{Arnaud2006}%
  \BibitemOpen
  \bibfield  {author} {\bibinfo {author} {\bibfnamefont {B.}~\bibnamefont
  {Arnaud}}, \bibinfo {author} {\bibfnamefont {S.}~\bibnamefont {Leb{\`e}gue}},
  \bibinfo {author} {\bibfnamefont {P.}~\bibnamefont {Rabiller}}, \ and\
  \bibinfo {author} {\bibfnamefont {M.}~\bibnamefont {Alouani}},\ }\href@noop
  {} {\bibfield  {journal} {\bibinfo  {journal} {Phys. Rev. Lett.}\ }\textbf
  {\bibinfo {volume} {96}},\ \bibinfo {pages} {026402} (\bibinfo {year}
  {2006})}\BibitemShut {NoStop}%
\bibitem [{\citenamefont {Wirtz}\ \emph {et~al.}(2008)\citenamefont {Wirtz},
  \citenamefont {Marini}, \citenamefont {Gr{\"u}ning}, \citenamefont
  {Attaccalite}, \citenamefont {Kresse},\ and\ \citenamefont
  {Rubio}}]{WirtzComment2008}%
  \BibitemOpen
  \bibfield  {author} {\bibinfo {author} {\bibfnamefont {L.}~\bibnamefont
  {Wirtz}}, \bibinfo {author} {\bibfnamefont {A.}~\bibnamefont {Marini}},
  \bibinfo {author} {\bibfnamefont {M.}~\bibnamefont {Gr{\"u}ning}}, \bibinfo
  {author} {\bibfnamefont {C.}~\bibnamefont {Attaccalite}}, \bibinfo {author}
  {\bibfnamefont {G.}~\bibnamefont {Kresse}}, \ and\ \bibinfo {author}
  {\bibfnamefont {A.}~\bibnamefont {Rubio}},\ }\href@noop {} {\bibfield
  {journal} {\bibinfo  {journal} {Phys. Rev. Lett.}\ }\textbf {\bibinfo
  {volume} {100}},\ \bibinfo {pages} {189701} (\bibinfo {year}
  {2008})}\BibitemShut {NoStop}%
\bibitem [{\citenamefont {Arnaud}\ \emph {et~al.}(2008)\citenamefont {Arnaud},
  \citenamefont {Leb{\`e}gue}, \citenamefont {Rabiller},\ and\ \citenamefont
  {Alouani}}]{ArnaudReply2008}%
  \BibitemOpen
  \bibfield  {author} {\bibinfo {author} {\bibfnamefont {B.}~\bibnamefont
  {Arnaud}}, \bibinfo {author} {\bibfnamefont {S.}~\bibnamefont {Leb{\`e}gue}},
  \bibinfo {author} {\bibfnamefont {P.}~\bibnamefont {Rabiller}}, \ and\
  \bibinfo {author} {\bibfnamefont {M.}~\bibnamefont {Alouani}},\ }\href@noop
  {} {\bibfield  {journal} {\bibinfo  {journal} {Phys. Rev. Lett.}\ }\textbf
  {\bibinfo {volume} {100}},\ \bibinfo {pages} {189702} (\bibinfo {year}
  {2008})}\BibitemShut {NoStop}%
\bibitem [{\citenamefont {Kim}\ \emph {et~al.}(2012)\citenamefont {Kim},
  \citenamefont {Hsu}, \citenamefont {Jia}, \citenamefont {Kim}, \citenamefont
  {Shi}, \citenamefont {Hofmann}, \citenamefont {Nezich}, \citenamefont
  {Rodriguez-Nieva}, \citenamefont {Dresselhaus}, \citenamefont {Palacios},\
  and\ \citenamefont {Kong}}]{Kim12}%
  \BibitemOpen
  \bibfield  {author} {\bibinfo {author} {\bibfnamefont {K.~K.}\ \bibnamefont
  {Kim}}, \bibinfo {author} {\bibfnamefont {A.}~\bibnamefont {Hsu}}, \bibinfo
  {author} {\bibfnamefont {X.}~\bibnamefont {Jia}}, \bibinfo {author}
  {\bibfnamefont {S.~M.}\ \bibnamefont {Kim}}, \bibinfo {author} {\bibfnamefont
  {Y.}~\bibnamefont {Shi}}, \bibinfo {author} {\bibfnamefont {M.}~\bibnamefont
  {Hofmann}}, \bibinfo {author} {\bibfnamefont {D.}~\bibnamefont {Nezich}},
  \bibinfo {author} {\bibfnamefont {J.~F.}\ \bibnamefont {Rodriguez-Nieva}},
  \bibinfo {author} {\bibfnamefont {M.}~\bibnamefont {Dresselhaus}}, \bibinfo
  {author} {\bibfnamefont {T.}~\bibnamefont {Palacios}}, \ and\ \bibinfo
  {author} {\bibfnamefont {J.}~\bibnamefont {Kong}},\ }\href {\doibase
  10.1021/nl203249a} {\bibfield  {journal} {\bibinfo  {journal} {Nano Lett.}\
  }\textbf {\bibinfo {volume} {12}},\ \bibinfo {pages} {161} (\bibinfo {year}
  {2012})}\BibitemShut {NoStop}%
\bibitem [{\citenamefont {Wang}\ \emph {et~al.}(2012)\citenamefont {Wang},
  \citenamefont {Pakdel}, \citenamefont {Zhi}, \citenamefont {Watanabe},
  \citenamefont {Sekiguchi}, \citenamefont {Golberg},\ and\ \citenamefont
  {Bando}}]{WangBN12}%
  \BibitemOpen
  \bibfield  {author} {\bibinfo {author} {\bibfnamefont {X.}~\bibnamefont
  {Wang}}, \bibinfo {author} {\bibfnamefont {A.}~\bibnamefont {Pakdel}},
  \bibinfo {author} {\bibfnamefont {C.}~\bibnamefont {Zhi}}, \bibinfo {author}
  {\bibfnamefont {K.}~\bibnamefont {Watanabe}}, \bibinfo {author}
  {\bibfnamefont {T.}~\bibnamefont {Sekiguchi}}, \bibinfo {author}
  {\bibfnamefont {D.}~\bibnamefont {Golberg}}, \ and\ \bibinfo {author}
  {\bibfnamefont {Y.}~\bibnamefont {Bando}},\ }\href@noop {} {\bibfield
  {journal} {\bibinfo  {journal} {J. Phys.: Condens. Matter}\ }\textbf
  {\bibinfo {volume} {24}},\ \bibinfo {pages} {314205} (\bibinfo {year}
  {2012})}\BibitemShut {NoStop}%
\bibitem [{\citenamefont {Ismach}\ \emph {et~al.}(2012)\citenamefont {Ismach},
  \citenamefont {Chou}, \citenamefont {Ferrer}, \citenamefont {Wu},
  \citenamefont {McDonnell}, \citenamefont {Floresca}, \citenamefont
  {Covacevich}, \citenamefont {Pope}, \citenamefont {Piner}, \citenamefont
  {Kim}, \citenamefont {Wallace}, \citenamefont {Colombo},\ and\ \citenamefont
  {Ruoff}}]{Ismach12}%
  \BibitemOpen
  \bibfield  {author} {\bibinfo {author} {\bibfnamefont {A.}~\bibnamefont
  {Ismach}}, \bibinfo {author} {\bibfnamefont {H.}~\bibnamefont {Chou}},
  \bibinfo {author} {\bibfnamefont {D.~A.}\ \bibnamefont {Ferrer}}, \bibinfo
  {author} {\bibfnamefont {Y.}~\bibnamefont {Wu}}, \bibinfo {author}
  {\bibfnamefont {S.}~\bibnamefont {McDonnell}}, \bibinfo {author}
  {\bibfnamefont {H.~C.}\ \bibnamefont {Floresca}}, \bibinfo {author}
  {\bibfnamefont {A.}~\bibnamefont {Covacevich}}, \bibinfo {author}
  {\bibfnamefont {C.}~\bibnamefont {Pope}}, \bibinfo {author} {\bibfnamefont
  {R.}~\bibnamefont {Piner}}, \bibinfo {author} {\bibfnamefont {M.~J.}\
  \bibnamefont {Kim}}, \bibinfo {author} {\bibfnamefont {R.~M.}\ \bibnamefont
  {Wallace}}, \bibinfo {author} {\bibfnamefont {L.}~\bibnamefont {Colombo}}, \
  and\ \bibinfo {author} {\bibfnamefont {R.~S.}\ \bibnamefont {Ruoff}},\ }\href
  {\doibase 10.1021/nn301940k} {\bibfield  {journal} {\bibinfo  {journal} {ACS
  Nano}\ }\textbf {\bibinfo {volume} {6}},\ \bibinfo {pages} {6378} (\bibinfo
  {year} {2012})}\BibitemShut {NoStop}%
\bibitem [{\citenamefont {Pakdel}\ \emph {et~al.}(2012)\citenamefont {Pakdel},
  \citenamefont {Wang}, \citenamefont {Zhi}, \citenamefont {Bando},
  \citenamefont {Watanabe}, \citenamefont {Sekiguchi}, \citenamefont
  {Nakayama},\ and\ \citenamefont {Golberg}}]{PakdelJMC12}%
  \BibitemOpen
  \bibfield  {author} {\bibinfo {author} {\bibfnamefont {A.}~\bibnamefont
  {Pakdel}}, \bibinfo {author} {\bibfnamefont {X.}~\bibnamefont {Wang}},
  \bibinfo {author} {\bibfnamefont {C.}~\bibnamefont {Zhi}}, \bibinfo {author}
  {\bibfnamefont {Y.}~\bibnamefont {Bando}}, \bibinfo {author} {\bibfnamefont
  {K.}~\bibnamefont {Watanabe}}, \bibinfo {author} {\bibfnamefont
  {T.}~\bibnamefont {Sekiguchi}}, \bibinfo {author} {\bibfnamefont
  {T.}~\bibnamefont {Nakayama}}, \ and\ \bibinfo {author} {\bibfnamefont
  {D.}~\bibnamefont {Golberg}},\ }\href {\doibase 10.1039/C2JM15109J}
  {\bibfield  {journal} {\bibinfo  {journal} {J. Mater. Chem.}\ }\textbf
  {\bibinfo {volume} {22}},\ \bibinfo {pages} {4818} (\bibinfo {year}
  {2012})}\BibitemShut {NoStop}%
\bibitem [{\citenamefont {Ci}\ \emph {et~al.}(2010)\citenamefont {Ci},
  \citenamefont {Song}, \citenamefont {Jin}, \citenamefont {Jariwala},
  \citenamefont {Wu}, \citenamefont {Li}, \citenamefont {Srivastava},
  \citenamefont {Wang}, \citenamefont {Storr}, \citenamefont {Balicas},
  \citenamefont {Lui},\ and\ \citenamefont {Ajayan}}]{Ci10}%
  \BibitemOpen
  \bibfield  {author} {\bibinfo {author} {\bibfnamefont {L.}~\bibnamefont
  {Ci}}, \bibinfo {author} {\bibfnamefont {L.}~\bibnamefont {Song}}, \bibinfo
  {author} {\bibfnamefont {C.}~\bibnamefont {Jin}}, \bibinfo {author}
  {\bibfnamefont {D.}~\bibnamefont {Jariwala}}, \bibinfo {author}
  {\bibfnamefont {D.}~\bibnamefont {Wu}}, \bibinfo {author} {\bibfnamefont
  {Y.}~\bibnamefont {Li}}, \bibinfo {author} {\bibfnamefont {A.}~\bibnamefont
  {Srivastava}}, \bibinfo {author} {\bibfnamefont {Z.}~\bibnamefont {Wang}},
  \bibinfo {author} {\bibfnamefont {K.}~\bibnamefont {Storr}}, \bibinfo
  {author} {\bibfnamefont {L.}~\bibnamefont {Balicas}}, \bibinfo {author}
  {\bibfnamefont {F.}~\bibnamefont {Lui}}, \ and\ \bibinfo {author}
  {\bibfnamefont {P.~M.}\ \bibnamefont {Ajayan}},\ }\href@noop {} {\bibfield
  {journal} {\bibinfo  {journal} {Nat. Mater.}\ }\textbf {\bibinfo {volume}
  {9}},\ \bibinfo {pages} {430} (\bibinfo {year} {2010})}\BibitemShut {NoStop}%
\bibitem [{\citenamefont {Lin}\ \emph {et~al.}(2009)\citenamefont {Lin},
  \citenamefont {Williams},\ and\ \citenamefont {Connell}}]{Lin2009}%
  \BibitemOpen
  \bibfield  {author} {\bibinfo {author} {\bibfnamefont {Y.}~\bibnamefont
  {Lin}}, \bibinfo {author} {\bibfnamefont {T.}~\bibnamefont {Williams}}, \
  and\ \bibinfo {author} {\bibfnamefont {J.}~\bibnamefont {Connell}},\
  }\href@noop {} {\bibfield  {journal} {\bibinfo  {journal} {The J. Phys. Chem.
  Lett.}\ }\textbf {\bibinfo {volume} {1}},\ \bibinfo {pages} {277} (\bibinfo
  {year} {2009})}\BibitemShut {NoStop}%
\bibitem [{\citenamefont {Gao}\ \emph {et~al.}(2012)\citenamefont {Gao},
  \citenamefont {Gao}, \citenamefont {Cannuccia}, \citenamefont
  {Taha-Tijerina}, \citenamefont {Balicas}, \citenamefont {Mathkar},
  \citenamefont {Narayanan}, \citenamefont {Liu}, \citenamefont {Gupta},
  \citenamefont {Peng}, \citenamefont {Yin}, \citenamefont {Rubio},\ and\
  \citenamefont {Ajayan}}]{Gao12}%
  \BibitemOpen
  \bibfield  {author} {\bibinfo {author} {\bibfnamefont {G.}~\bibnamefont
  {Gao}}, \bibinfo {author} {\bibfnamefont {W.}~\bibnamefont {Gao}}, \bibinfo
  {author} {\bibfnamefont {E.}~\bibnamefont {Cannuccia}}, \bibinfo {author}
  {\bibfnamefont {J.}~\bibnamefont {Taha-Tijerina}}, \bibinfo {author}
  {\bibfnamefont {L.}~\bibnamefont {Balicas}}, \bibinfo {author} {\bibfnamefont
  {A.}~\bibnamefont {Mathkar}}, \bibinfo {author} {\bibfnamefont {T.~N.}\
  \bibnamefont {Narayanan}}, \bibinfo {author} {\bibfnamefont {Z.}~\bibnamefont
  {Liu}}, \bibinfo {author} {\bibfnamefont {B.~K.}\ \bibnamefont {Gupta}},
  \bibinfo {author} {\bibfnamefont {J.}~\bibnamefont {Peng}}, \bibinfo {author}
  {\bibfnamefont {Y.}~\bibnamefont {Yin}}, \bibinfo {author} {\bibfnamefont
  {A.}~\bibnamefont {Rubio}}, \ and\ \bibinfo {author} {\bibfnamefont {P.~M.}\
  \bibnamefont {Ajayan}},\ }\href {\doibase 10.1021/nl301061b} {\bibfield
  {journal} {\bibinfo  {journal} {Nano Lett.}\ }\textbf {\bibinfo {volume}
  {12}},\ \bibinfo {pages} {3518} (\bibinfo {year} {2012})}\BibitemShut
  {NoStop}%
\bibitem [{\citenamefont {Li}\ \emph {et~al.}(2012)\citenamefont {Li},
  \citenamefont {Chen}, \citenamefont {Cheng}, \citenamefont {Lin},
  \citenamefont {Chou},\ and\ \citenamefont {Peng}}]{Li12}%
  \BibitemOpen
  \bibfield  {author} {\bibinfo {author} {\bibfnamefont {L.~H.}\ \bibnamefont
  {Li}}, \bibinfo {author} {\bibfnamefont {Y.}~\bibnamefont {Chen}}, \bibinfo
  {author} {\bibfnamefont {B.-M.}\ \bibnamefont {Cheng}}, \bibinfo {author}
  {\bibfnamefont {M.-Y.}\ \bibnamefont {Lin}}, \bibinfo {author} {\bibfnamefont
  {S.-L.}\ \bibnamefont {Chou}}, \ and\ \bibinfo {author} {\bibfnamefont
  {Y.-C.}\ \bibnamefont {Peng}},\ }\href {\doibase 10.1063/1.4731203}
  {\bibfield  {journal} {\bibinfo  {journal} {Appl. Phys. Lett.}\ }\textbf
  {\bibinfo {volume} {100}},\ \bibinfo {eid} {261108} (\bibinfo {year}
  {2012})}\BibitemShut {NoStop}%
\bibitem [{\citenamefont {Taniguchi}\ and\ \citenamefont
  {Watanabe}(2007)}]{Taniguchi2007}%
  \BibitemOpen
  \bibfield  {author} {\bibinfo {author} {\bibfnamefont {T.}~\bibnamefont
  {Taniguchi}}\ and\ \bibinfo {author} {\bibfnamefont {K.}~\bibnamefont
  {Watanabe}},\ }\href@noop {} {\bibfield  {journal} {\bibinfo  {journal} {J.
  Cryst. Growth}\ }\textbf {\bibinfo {volume} {303}},\ \bibinfo {pages} {525}
  (\bibinfo {year} {2007})}\BibitemShut {NoStop}%
\bibitem [{\citenamefont {Novoselov}\ \emph {et~al.}(2004)\citenamefont
  {Novoselov}, \citenamefont {Geim}, \citenamefont {Morozov}, \citenamefont
  {Jiang}, \citenamefont {Zhang}, \citenamefont {Dubonos}, \citenamefont
  {Grigorieva},\ and\ \citenamefont {Firsov}}]{Novoselov2004}%
  \BibitemOpen
  \bibfield  {author} {\bibinfo {author} {\bibfnamefont {K.}~\bibnamefont
  {Novoselov}}, \bibinfo {author} {\bibfnamefont {A.}~\bibnamefont {Geim}},
  \bibinfo {author} {\bibfnamefont {S.}~\bibnamefont {Morozov}}, \bibinfo
  {author} {\bibfnamefont {D.}~\bibnamefont {Jiang}}, \bibinfo {author}
  {\bibfnamefont {Y.}~\bibnamefont {Zhang}}, \bibinfo {author} {\bibfnamefont
  {S.}~\bibnamefont {Dubonos}}, \bibinfo {author} {\bibfnamefont
  {I.}~\bibnamefont {Grigorieva}}, \ and\ \bibinfo {author} {\bibfnamefont
  {A.}~\bibnamefont {Firsov}},\ }\href@noop {} {\bibfield  {journal} {\bibinfo
  {journal} {Science}\ }\textbf {\bibinfo {volume} {306}},\ \bibinfo {pages}
  {666} (\bibinfo {year} {2004})}\BibitemShut {NoStop}%
\bibitem [{\citenamefont {Gorbachev}\ \emph {et~al.}(2011)\citenamefont
  {Gorbachev}, \citenamefont {Riaz}, \citenamefont {Nair}, \citenamefont
  {Jalil}, \citenamefont {Britnell}, \citenamefont {Belle}, \citenamefont
  {Hill}, \citenamefont {Novoselov}, \citenamefont {Watanabe}, \citenamefont
  {Taniguchi}, \citenamefont {Geim},\ and\ \citenamefont
  {Blake}}]{Gorbachev11}%
  \BibitemOpen
  \bibfield  {author} {\bibinfo {author} {\bibfnamefont {R.~V.}\ \bibnamefont
  {Gorbachev}}, \bibinfo {author} {\bibfnamefont {I.}~\bibnamefont {Riaz}},
  \bibinfo {author} {\bibfnamefont {R.~R.}\ \bibnamefont {Nair}}, \bibinfo
  {author} {\bibfnamefont {R.}~\bibnamefont {Jalil}}, \bibinfo {author}
  {\bibfnamefont {L.}~\bibnamefont {Britnell}}, \bibinfo {author}
  {\bibfnamefont {B.~D.}\ \bibnamefont {Belle}}, \bibinfo {author}
  {\bibfnamefont {E.~W.}\ \bibnamefont {Hill}}, \bibinfo {author}
  {\bibfnamefont {K.~S.}\ \bibnamefont {Novoselov}}, \bibinfo {author}
  {\bibfnamefont {K.}~\bibnamefont {Watanabe}}, \bibinfo {author}
  {\bibfnamefont {T.}~\bibnamefont {Taniguchi}}, \bibinfo {author}
  {\bibfnamefont {A.~K.}\ \bibnamefont {Geim}}, \ and\ \bibinfo {author}
  {\bibfnamefont {P.}~\bibnamefont {Blake}},\ }\href {\doibase
  10.1002/smll.201001628} {\bibfield  {journal} {\bibinfo  {journal} {Small}\
  }\textbf {\bibinfo {volume} {7}},\ \bibinfo {pages} {465} (\bibinfo {year}
  {2011})}\BibitemShut {NoStop}%
\bibitem [{\citenamefont {Betz}(2012)}]{BetzPhD}%
  \BibitemOpen
  \bibfield  {author} {\bibinfo {author} {\bibfnamefont {A.}~\bibnamefont
  {Betz}},\ }\emph {\bibinfo {title} {RF dynamics and noise of graphene
  microwave devices}},\ \href@noop {} {Ph.D. thesis},\ \bibinfo  {school}
  {Paris VI University} (\bibinfo {year} {2012})\BibitemShut {NoStop}%
\bibitem [{\citenamefont {Nagashio}\ \emph {et~al.}(2011)\citenamefont
  {Nagashio}, \citenamefont {Yamashita}, \citenamefont {Nishimura},
  \citenamefont {Kita},\ and\ \citenamefont {Toriumi}}]{Nagashio11}%
  \BibitemOpen
  \bibfield  {author} {\bibinfo {author} {\bibfnamefont {K.}~\bibnamefont
  {Nagashio}}, \bibinfo {author} {\bibfnamefont {T.}~\bibnamefont {Yamashita}},
  \bibinfo {author} {\bibfnamefont {T.}~\bibnamefont {Nishimura}}, \bibinfo
  {author} {\bibfnamefont {K.}~\bibnamefont {Kita}}, \ and\ \bibinfo {author}
  {\bibfnamefont {A.}~\bibnamefont {Toriumi}},\ }\href {\doibase
  10.1063/1.3611394} {\bibfield  {journal} {\bibinfo  {journal} {J. Appl.
  Phys.}\ }\textbf {\bibinfo {volume} {110}},\ \bibinfo {pages} {024513}
  (\bibinfo {year} {2011})}\BibitemShut {NoStop}%
\bibitem [{\citenamefont {Ishigami}\ \emph {et~al.}(2007)\citenamefont
  {Ishigami}, \citenamefont {Chen}, \citenamefont {Cullen}, \citenamefont
  {Fuhrer},\ and\ \citenamefont {Williams}}]{Ishigami2007}%
  \BibitemOpen
  \bibfield  {author} {\bibinfo {author} {\bibfnamefont {M.}~\bibnamefont
  {Ishigami}}, \bibinfo {author} {\bibfnamefont {J.~H.}\ \bibnamefont {Chen}},
  \bibinfo {author} {\bibfnamefont {W.~G.}\ \bibnamefont {Cullen}}, \bibinfo
  {author} {\bibfnamefont {M.~S.}\ \bibnamefont {Fuhrer}}, \ and\ \bibinfo
  {author} {\bibfnamefont {E.~D.}\ \bibnamefont {Williams}},\ }\href {\doibase
  10.1021/nl070613a} {\bibfield  {journal} {\bibinfo  {journal} {Nano Lett.}\
  }\textbf {\bibinfo {volume} {7}},\ \bibinfo {pages} {1643} (\bibinfo {year}
  {2007})}\BibitemShut {NoStop}%
\bibitem [{\citenamefont {Watanabe}\ \emph {et~al.}(2006)\citenamefont
  {Watanabe}, \citenamefont {Taniguchi}, \citenamefont {Kuroda},\ and\
  \citenamefont {Kanda}}]{WatanabeAPL2006}%
  \BibitemOpen
  \bibfield  {author} {\bibinfo {author} {\bibfnamefont {K.}~\bibnamefont
  {Watanabe}}, \bibinfo {author} {\bibfnamefont {T.}~\bibnamefont {Taniguchi}},
  \bibinfo {author} {\bibfnamefont {T.}~\bibnamefont {Kuroda}}, \ and\ \bibinfo
  {author} {\bibfnamefont {H.}~\bibnamefont {Kanda}},\ }\href@noop {}
  {\bibfield  {journal} {\bibinfo  {journal} {Appl. Phys. Lett.}\ }\textbf
  {\bibinfo {volume} {89}},\ \bibinfo {pages} {141902} (\bibinfo {year}
  {2006})}\BibitemShut {NoStop}%
  \bibitem [{\citenamefont {Kubota}\ \emph {et~al.}(2007)\citenamefont {Kubota},
  \citenamefont {Watanabe}, \citenamefont {Tsuda},\ and\ \citenamefont
  {Taniguchi}}]{Kubota2007}%
  \BibitemOpen
  \bibfield  {author} {\bibinfo {author} {\bibfnamefont {Y.}~\bibnamefont
  {Kubota}}, \bibinfo {author} {\bibfnamefont {K.}~\bibnamefont {Watanabe}},
  \bibinfo {author} {\bibfnamefont {O.}~\bibnamefont {Tsuda}}, \ and\ \bibinfo
  {author} {\bibfnamefont {T.}~\bibnamefont {Taniguchi}},\ }\href@noop {}
  {\bibfield  {journal} {\bibinfo  {journal} {Science}\ }\textbf {\bibinfo
  {volume} {317}},\ \bibinfo {pages} {932} (\bibinfo {year}
  {2007})}\BibitemShut {NoStop}%
\bibitem [{\citenamefont {Barjon}\ \emph {et~al.}(2011)\citenamefont {Barjon},
  \citenamefont {Tillocher}, \citenamefont {Habka}, \citenamefont {Brinza},
  \citenamefont {Achard}, \citenamefont {Issaoui}, \citenamefont {Silva},
  \citenamefont {Mer},\ and\ \citenamefont {Bergonzo}}]{Barjon11}%
  \BibitemOpen
  \bibfield  {author} {\bibinfo {author} {\bibfnamefont {J.}~\bibnamefont
  {Barjon}}, \bibinfo {author} {\bibfnamefont {T.}~\bibnamefont {Tillocher}},
  \bibinfo {author} {\bibfnamefont {N.}~\bibnamefont {Habka}}, \bibinfo
  {author} {\bibfnamefont {O.}~\bibnamefont {Brinza}}, \bibinfo {author}
  {\bibfnamefont {J.}~\bibnamefont {Achard}}, \bibinfo {author} {\bibfnamefont
  {R.}~\bibnamefont {Issaoui}}, \bibinfo {author} {\bibfnamefont
  {F.}~\bibnamefont {Silva}}, \bibinfo {author} {\bibfnamefont
  {C.}~\bibnamefont {Mer}}, \ and\ \bibinfo {author} {\bibfnamefont
  {P.}~\bibnamefont {Bergonzo}},\ }\href {\doibase 10.1103/PhysRevB.83.073201}
  {\bibfield  {journal} {\bibinfo  {journal} {Phys. Rev. B}\ }\textbf {\bibinfo
  {volume} {83}},\ \bibinfo {pages} {073201} (\bibinfo {year}
  {2011})}\BibitemShut {NoStop}%
\bibitem [{\citenamefont {Serrano}\ \emph {et~al.}(2007)\citenamefont
  {Serrano}, \citenamefont {Bosak}, \citenamefont {Arenal}, \citenamefont
  {Krisch}, \citenamefont {Watanabe}, \citenamefont {Taniguchi}, \citenamefont
  {Kanda}, \citenamefont {Rubio},\ and\ \citenamefont {Wirtz}}]{Serrano2007}%
  \BibitemOpen
  \bibfield  {author} {\bibinfo {author} {\bibfnamefont {J.}~\bibnamefont
  {Serrano}}, \bibinfo {author} {\bibfnamefont {A.}~\bibnamefont {Bosak}},
  \bibinfo {author} {\bibfnamefont {R.}~\bibnamefont {Arenal}}, \bibinfo
  {author} {\bibfnamefont {M.}~\bibnamefont {Krisch}}, \bibinfo {author}
  {\bibfnamefont {K.}~\bibnamefont {Watanabe}}, \bibinfo {author}
  {\bibfnamefont {T.}~\bibnamefont {Taniguchi}}, \bibinfo {author}
  {\bibfnamefont {H.}~\bibnamefont {Kanda}}, \bibinfo {author} {\bibfnamefont
  {A.}~\bibnamefont {Rubio}}, \ and\ \bibinfo {author} {\bibfnamefont
  {L.}~\bibnamefont {Wirtz}},\ }\href@noop {} {\bibfield  {journal} {\bibinfo
  {journal} {Phys. Rev. Lett.}\ }\textbf {\bibinfo {volume} {98}},\ \bibinfo
  {pages} {095503} (\bibinfo {year} {2007})}\BibitemShut {NoStop}%
\bibitem [{\citenamefont {Silly}\ \emph {et~al.}(2007)\citenamefont {Silly},
  \citenamefont {Jaffrennou}, \citenamefont {Barjon}, \citenamefont {Lauret},
  \citenamefont {Ducastelle}, \citenamefont {Loiseau}, \citenamefont
  {Obraztsova}, \citenamefont {Attal-Tr{\'e}tout},\ and\ \citenamefont
  {Rosencher}}]{Silly2007}%
  \BibitemOpen
  \bibfield  {author} {\bibinfo {author} {\bibfnamefont {M.~G.}\ \bibnamefont
  {Silly}}, \bibinfo {author} {\bibfnamefont {P.}~\bibnamefont {Jaffrennou}},
  \bibinfo {author} {\bibfnamefont {J.}~\bibnamefont {Barjon}}, \bibinfo
  {author} {\bibfnamefont {J.-S.}\ \bibnamefont {Lauret}}, \bibinfo {author}
  {\bibfnamefont {F.}~\bibnamefont {Ducastelle}}, \bibinfo {author}
  {\bibfnamefont {A.}~\bibnamefont {Loiseau}}, \bibinfo {author} {\bibfnamefont
  {E.}~\bibnamefont {Obraztsova}}, \bibinfo {author} {\bibfnamefont
  {B.}~\bibnamefont {Attal-Tr{\'e}tout}}, \ and\ \bibinfo {author}
  {\bibfnamefont {E.}~\bibnamefont {Rosencher}},\ }\href@noop {} {\bibfield
  {journal} {\bibinfo  {journal} {Phys. Rev. B}\ }\textbf {\bibinfo {volume}
  {75}},\ \bibinfo {pages} {085205} (\bibinfo {year} {2007})}\BibitemShut
  {NoStop}%
\bibitem [{\citenamefont {Alem}\ \emph {et~al.}(2012)\citenamefont {Alem},
  \citenamefont {Ramasse}, \citenamefont {Seabourne}, \citenamefont {Yazyev},
  \citenamefont {Erickson}, \citenamefont {Sarahan}, \citenamefont
  {Kisielowski}, \citenamefont {Scott}, \citenamefont {Louie},\ and\
  \citenamefont {Zettl}}]{Alem12}%
  \BibitemOpen
  \bibfield  {author} {\bibinfo {author} {\bibfnamefont {N.}~\bibnamefont
  {Alem}}, \bibinfo {author} {\bibfnamefont {Q.~M.}\ \bibnamefont {Ramasse}},
  \bibinfo {author} {\bibfnamefont {C.~R.}\ \bibnamefont {Seabourne}}, \bibinfo
  {author} {\bibfnamefont {O.~V.}\ \bibnamefont {Yazyev}}, \bibinfo {author}
  {\bibfnamefont {K.}~\bibnamefont {Erickson}}, \bibinfo {author}
  {\bibfnamefont {M.~C.}\ \bibnamefont {Sarahan}}, \bibinfo {author}
  {\bibfnamefont {C.}~\bibnamefont {Kisielowski}}, \bibinfo {author}
  {\bibfnamefont {A.~J.}\ \bibnamefont {Scott}}, \bibinfo {author}
  {\bibfnamefont {S.~G.}\ \bibnamefont {Louie}}, \ and\ \bibinfo {author}
  {\bibfnamefont {A.}~\bibnamefont {Zettl}},\ }\href {\doibase
  10.1103/PhysRevLett.109.205502} {\bibfield  {journal} {\bibinfo  {journal}
  {Phys. Rev. Lett.}\ }\textbf {\bibinfo {volume} {109}},\ \bibinfo {pages}
  {205502} (\bibinfo {year} {2012})}\BibitemShut {NoStop}%
\bibitem [{\citenamefont {Jin}\ \emph {et~al.}(2009)\citenamefont {Jin},
  \citenamefont {Lin}, \citenamefont {Suenaga},\ and\ \citenamefont
  {Iijima}}]{Jin2009}%
  \BibitemOpen
  \bibfield  {author} {\bibinfo {author} {\bibfnamefont {C.}~\bibnamefont
  {Jin}}, \bibinfo {author} {\bibfnamefont {F.}~\bibnamefont {Lin}}, \bibinfo
  {author} {\bibfnamefont {K.}~\bibnamefont {Suenaga}}, \ and\ \bibinfo
  {author} {\bibfnamefont {S.}~\bibnamefont {Iijima}},\ }\href@noop {}
  {\bibfield  {journal} {\bibinfo  {journal} {Phys. Rev. Lett.}\ }\textbf
  {\bibinfo {volume} {102}},\ \bibinfo {pages} {195505} (\bibinfo {year}
  {2009})}\BibitemShut {NoStop}%
\bibitem [{\citenamefont {Suenaga}\ \emph {et~al.}(2012)\citenamefont
  {Suenaga}, \citenamefont {Kobayashi},\ and\ \citenamefont
  {Koshino}}]{Suenaga12}%
  \BibitemOpen
  \bibfield  {author} {\bibinfo {author} {\bibfnamefont {K.}~\bibnamefont
  {Suenaga}}, \bibinfo {author} {\bibfnamefont {H.}~\bibnamefont {Kobayashi}},
  \ and\ \bibinfo {author} {\bibfnamefont {M.}~\bibnamefont {Koshino}},\ }\href
  {\doibase 10.1103/PhysRevLett.108.075501} {\bibfield  {journal} {\bibinfo
  {journal} {Phys. Rev. Lett.}\ }\textbf {\bibinfo {volume} {108}},\ \bibinfo
  {pages} {075501} (\bibinfo {year} {2012})}\BibitemShut {NoStop}%
\bibitem [{\citenamefont {Pan}\ \emph {et~al.}(2012)\citenamefont {Pan},
  \citenamefont {Nair}, \citenamefont {Bangert}, \citenamefont {Ramasse},
  \citenamefont {Jalil}, \citenamefont {Zan}, \citenamefont {Seabourne},\ and\
  \citenamefont {Scott}}]{Pan12}%
  \BibitemOpen
  \bibfield  {author} {\bibinfo {author} {\bibfnamefont {C.~T.}\ \bibnamefont
  {Pan}}, \bibinfo {author} {\bibfnamefont {R.~R.}\ \bibnamefont {Nair}},
  \bibinfo {author} {\bibfnamefont {U.}~\bibnamefont {Bangert}}, \bibinfo
  {author} {\bibfnamefont {Q.}~\bibnamefont {Ramasse}}, \bibinfo {author}
  {\bibfnamefont {R.}~\bibnamefont {Jalil}}, \bibinfo {author} {\bibfnamefont
  {R.}~\bibnamefont {Zan}}, \bibinfo {author} {\bibfnamefont {C.~R.}\
  \bibnamefont {Seabourne}}, \ and\ \bibinfo {author} {\bibfnamefont {A.~J.}\
  \bibnamefont {Scott}},\ }\href {\doibase 10.1103/PhysRevB.85.045440}
  {\bibfield  {journal} {\bibinfo  {journal} {Phys. Rev. B}\ }\textbf {\bibinfo
  {volume} {85}},\ \bibinfo {pages} {045440} (\bibinfo {year}
  {2012})}\BibitemShut {NoStop}%
\bibitem [{\citenamefont {Kalceff}\ \emph {et~al.}(1996)\citenamefont
  {Kalceff}, \citenamefont {Phillips},\ and\ \citenamefont
  {Moon}}]{Kalceff1996}%
  \BibitemOpen
  \bibfield  {author} {\bibinfo {author} {\bibfnamefont {M.~A.~S.}\
  \bibnamefont {Kalceff}}, \bibinfo {author} {\bibfnamefont {M.~R.}\
  \bibnamefont {Phillips}}, \ and\ \bibinfo {author} {\bibfnamefont {A.~R.}\
  \bibnamefont {Moon}},\ }\href {\doibase 10.1063/1.363379} {\bibfield
  {journal} {\bibinfo  {journal} {J. Appl. Phys.}\ }\textbf {\bibinfo {volume}
  {80}},\ \bibinfo {pages} {4308} (\bibinfo {year} {1996})}\BibitemShut
  {NoStop}%
\bibitem [{\citenamefont {Demichel}\ \emph {et~al.}(2010)\citenamefont
  {Demichel}, \citenamefont {Heiss}, \citenamefont {Bleuse}, \citenamefont
  {Mariette},\ and\ \citenamefont {Fontcuberta~i Morral}}]{Demichel10}%
  \BibitemOpen
  \bibfield  {author} {\bibinfo {author} {\bibfnamefont {O.}~\bibnamefont
  {Demichel}}, \bibinfo {author} {\bibfnamefont {M.}~\bibnamefont {Heiss}},
  \bibinfo {author} {\bibfnamefont {J.}~\bibnamefont {Bleuse}}, \bibinfo
  {author} {\bibfnamefont {H.}~\bibnamefont {Mariette}}, \ and\ \bibinfo
  {author} {\bibfnamefont {A.}~\bibnamefont {Fontcuberta~i Morral}},\ }\href
  {\doibase 10.1063/1.3519980} {\bibfield  {journal} {\bibinfo  {journal}
  {Appl. Phys. Lett.}\ }\textbf {\bibinfo {volume} {97}},\ \bibinfo {pages}
  {201907} (\bibinfo {year} {2010})}\BibitemShut {NoStop}%
\bibitem [{\citenamefont {Calarco}\ \emph {et~al.}(2011)\citenamefont
  {Calarco}, \citenamefont {Stoica}, \citenamefont {Brandt},\ and\
  \citenamefont {Geelhaar}}]{Calarco11}%
  \BibitemOpen
  \bibfield  {author} {\bibinfo {author} {\bibfnamefont {R.}~\bibnamefont
  {Calarco}}, \bibinfo {author} {\bibfnamefont {T.}~\bibnamefont {Stoica}},
  \bibinfo {author} {\bibfnamefont {O.}~\bibnamefont {Brandt}}, \ and\ \bibinfo
  {author} {\bibfnamefont {L.}~\bibnamefont {Geelhaar}},\ }\href {\doibase
  10.1557/jmr.2011.211} {\bibfield  {journal} {\bibinfo  {journal} {J. Mater.
  Res.}\ }\textbf {\bibinfo {volume} {26}},\ \bibinfo {pages} {2157} (\bibinfo
  {year} {2011})}\BibitemShut {NoStop}%
\bibitem [{\citenamefont {Hines}\ and\ \citenamefont
  {Guyot-Sionnest}(1996)}]{Hines1996}%
  \BibitemOpen
  \bibfield  {author} {\bibinfo {author} {\bibfnamefont {M.~A.}\ \bibnamefont
  {Hines}}\ and\ \bibinfo {author} {\bibfnamefont {P.}~\bibnamefont
  {Guyot-Sionnest}},\ }\href {\doibase 10.1021/jp9530562} {\bibfield  {journal}
  {\bibinfo  {journal} {J. Phys. Chem.}\ }\textbf {\bibinfo {volume} {100}},\
  \bibinfo {pages} {468} (\bibinfo {year} {1996})}\BibitemShut {NoStop}%
\bibitem [{\citenamefont {Peng}\ \emph {et~al.}(1997)\citenamefont {Peng},
  \citenamefont {Schlamp}, \citenamefont {Kadavanich},\ and\ \citenamefont
  {Alivisatos}}]{Peng1997}%
  \BibitemOpen
  \bibfield  {author} {\bibinfo {author} {\bibfnamefont {X.}~\bibnamefont
  {Peng}}, \bibinfo {author} {\bibfnamefont {M.~C.}\ \bibnamefont {Schlamp}},
  \bibinfo {author} {\bibfnamefont {A.~V.}\ \bibnamefont {Kadavanich}}, \ and\
  \bibinfo {author} {\bibfnamefont {A.~P.}\ \bibnamefont {Alivisatos}},\ }\href
  {\doibase 10.1021/ja970754m} {\bibfield  {journal} {\bibinfo  {journal} {J.
  Am. Chem. Soc.}\ }\textbf {\bibinfo {volume} {119}},\ \bibinfo {pages} {7019}
  (\bibinfo {year} {1997})}\BibitemShut {NoStop}%
\bibitem [{\citenamefont {Couto}\ \emph {et~al.}(2012)\citenamefont {Couto},
  \citenamefont {Sercombe}, \citenamefont {Puebla}, \citenamefont {Otubo},
  \citenamefont {Luxmoore}, \citenamefont {Sich}, \citenamefont {Elliott},
  \citenamefont {Chekhovich}, \citenamefont {Wilson}, \citenamefont {Skolnick},
  \citenamefont {Liu},\ and\ \citenamefont {Tartakovskii}}]{Couto12}%
  \BibitemOpen
  \bibfield  {author} {\bibinfo {author} {\bibfnamefont {O.~D.~D.}\
  \bibnamefont {Couto}}, \bibinfo {author} {\bibfnamefont {D.}~\bibnamefont
  {Sercombe}}, \bibinfo {author} {\bibfnamefont {J.}~\bibnamefont {Puebla}},
  \bibinfo {author} {\bibfnamefont {L.}~\bibnamefont {Otubo}}, \bibinfo
  {author} {\bibfnamefont {I.~J.}\ \bibnamefont {Luxmoore}}, \bibinfo {author}
  {\bibfnamefont {M.}~\bibnamefont {Sich}}, \bibinfo {author} {\bibfnamefont
  {T.~J.}\ \bibnamefont {Elliott}}, \bibinfo {author} {\bibfnamefont {E.~A.}\
  \bibnamefont {Chekhovich}}, \bibinfo {author} {\bibfnamefont {L.~R.}\
  \bibnamefont {Wilson}}, \bibinfo {author} {\bibfnamefont {M.~S.}\
  \bibnamefont {Skolnick}}, \bibinfo {author} {\bibfnamefont {H.~Y.}\
  \bibnamefont {Liu}}, \ and\ \bibinfo {author} {\bibfnamefont {A.~I.}\
  \bibnamefont {Tartakovskii}},\ }\href {\doibase 10.1021/nl302490y} {\bibfield
   {journal} {\bibinfo  {journal} {Nano Lett.}\ }\textbf {\bibinfo {volume}
  {12}},\ \bibinfo {pages} {5269} (\bibinfo {year} {2012})}\BibitemShut
  {NoStop}%
\bibitem [{\citenamefont {Titova}\ \emph {et~al.}(2006)\citenamefont {Titova},
  \citenamefont {Hoang}, \citenamefont {Jackson}, \citenamefont {Smith},
  \citenamefont {Yarrison-Rice}, \citenamefont {Kim}, \citenamefont {Joyce},
  \citenamefont {Tan},\ and\ \citenamefont {Jagadish}}]{Titova2006}%
  \BibitemOpen
  \bibfield  {author} {\bibinfo {author} {\bibfnamefont {L.~V.}\ \bibnamefont
  {Titova}}, \bibinfo {author} {\bibfnamefont {T.~B.}\ \bibnamefont {Hoang}},
  \bibinfo {author} {\bibfnamefont {H.~E.}\ \bibnamefont {Jackson}}, \bibinfo
  {author} {\bibfnamefont {L.~M.}\ \bibnamefont {Smith}}, \bibinfo {author}
  {\bibfnamefont {J.~M.}\ \bibnamefont {Yarrison-Rice}}, \bibinfo {author}
  {\bibfnamefont {Y.}~\bibnamefont {Kim}}, \bibinfo {author} {\bibfnamefont
  {H.~J.}\ \bibnamefont {Joyce}}, \bibinfo {author} {\bibfnamefont {H.~H.}\
  \bibnamefont {Tan}}, \ and\ \bibinfo {author} {\bibfnamefont
  {C.}~\bibnamefont {Jagadish}},\ }\href {\doibase 10.1063/1.2364885}
  {\bibfield  {journal} {\bibinfo  {journal} {Appl. Phys. Lett.}\ }\textbf
  {\bibinfo {volume} {89}},\ \bibinfo {eid} {173126} (\bibinfo {year}
  {2006})}\BibitemShut {NoStop}%
\bibitem [{\citenamefont {Wirtz}\ \emph {et~al.}(2006)\citenamefont {Wirtz},
  \citenamefont {Marini},\ and\ \citenamefont {Rubio}}]{Wirtz2006}%
  \BibitemOpen
  \bibfield  {author} {\bibinfo {author} {\bibfnamefont {L.}~\bibnamefont
  {Wirtz}}, \bibinfo {author} {\bibfnamefont {A.}~\bibnamefont {Marini}}, \
  and\ \bibinfo {author} {\bibfnamefont {A.}~\bibnamefont {Rubio}},\
  }\href@noop {} {\bibfield  {journal} {\bibinfo  {journal} {Phys. Rev. Lett.}\
  }\textbf {\bibinfo {volume} {96}},\ \bibinfo {pages} {126104} (\bibinfo
  {year} {2006})}\BibitemShut {NoStop}%
\bibitem [{\citenamefont {Park}\ \emph {et~al.}(2006)\citenamefont {Park},
  \citenamefont {Spataru},\ and\ \citenamefont {Louie}}]{ParkBN2006}%
  \BibitemOpen
  \bibfield  {author} {\bibinfo {author} {\bibfnamefont {C.-H.}\ \bibnamefont
  {Park}}, \bibinfo {author} {\bibfnamefont {C.~D.}\ \bibnamefont {Spataru}}, \
  and\ \bibinfo {author} {\bibfnamefont {S.~G.}\ \bibnamefont {Louie}},\
  }\href@noop {} {\bibfield  {journal} {\bibinfo  {journal} {Phys. Rev. Lett.}\
  }\textbf {\bibinfo {volume} {96}},\ \bibinfo {pages} {126105} (\bibinfo
  {year} {2006})}\BibitemShut {NoStop}%
\bibitem [{\citenamefont {Wang}\ \emph {et~al.}(2011)\citenamefont {Wang},
  \citenamefont {Chen},\ and\ \citenamefont {Wang}}]{Wang11}%
  \BibitemOpen
  \bibfield  {author} {\bibinfo {author} {\bibfnamefont {S.}~\bibnamefont
  {Wang}}, \bibinfo {author} {\bibfnamefont {Q.}~\bibnamefont {Chen}}, \ and\
  \bibinfo {author} {\bibfnamefont {J.}~\bibnamefont {Wang}},\ }\href {\doibase
  10.1063/1.3625922} {\bibfield  {journal} {\bibinfo  {journal} {Appl. Phys.
  Lett.}\ }\textbf {\bibinfo {volume} {99}},\ \bibinfo {eid} {063114} (\bibinfo
  {year} {2011})}\BibitemShut {NoStop}%
\end{thebibliography}
\end{document}